\documentclass[usenatbib,useAMS]{mnras}

\usepackage{graphicx}	
\usepackage{amsmath}	
\usepackage{amssymb}	
\usepackage{multicol}        
\usepackage{bm}		
\usepackage{pdflscape}	

\usepackage{setspace}\usepackage{threeparttable}

\usepackage[T1]{fontenc}
\usepackage{ae,aecompl}

\usepackage{newtxtext,newtxmath}

\title{Turbulence-Level Dependence of Cosmic-Ray Parallel Diffusion}

\author[Reichherzer et al.]{P.~Reichherzer$^{1,2,3}$\thanks{Contact e-mail: \href{mailto:patrick.reichherzer@rub.de}{patrick.reichherzer@rub.de}}, J.~Becker Tjus$^{2,3}$, E.G.~Zweibel$^{4,5}$, L.~Merten$^{2,3,6}$, and M.J.~Pueschel$^7$\\
$^1$ IRFU, CEA, Université Paris-Saclay, F-91191 Gif-sur-Yvette, France\\
$^2$ Theoretical Physics IV: Plasma-Astroparticle Physics, Faculty for Physics \& Astronomy, Ruhr-Universit\"at Bochum, 44780 Bochum, Germany\\
$^3$ Ruhr Astroparticle And Plasma Physics Center (RAPP Center)\\
$^4$ Department of Astronomy, University of Wisconsin-Madison, Madison, WI 53706, U.S.A.\\
$^5$ Department of Physics, University of Wisconsin-Madison, Madison, WI 53706, U.S.A.\\
$^6$ Institute for Astro- \& Particle Physics, University of Innsbruck, 6020 Innsbruck, Austria\\
$^7$ Institute for Fusion Studies, University of Texas at Austin, Austin, TX 78712, U.S.A.}

\date{Last updated 2015 May 22; in original form 2013 September 5}

\pubyear{2019}

\begin{document}
\label{firstpage}
\pagerange{\pageref{firstpage}--\pageref{lastpage}}
\maketitle

\begin{abstract}
Understanding the transport of energetic cosmic rays belongs to the most challenging topics in astrophysics. Diffusion due to scattering by electromagnetic fluctuations is a key process in cosmic-ray transport. The transition from a ballistic to a diffusive-propagation regime is presented in direct numerical calculations of diffusion coefficients for homogeneous magnetic field lines subject to turbulent perturbations.
Simulation results are compared with theoretical derivations of the parallel diffusion coefficient's dependencies on the energy and the fluctuation amplitudes in the limit of weak turbulence. The present study shows that the widely-used extrapolation of the energy scaling for the parallel diffusion coefficient to high turbulence levels predicted by quasi-linear theory does not provide a universally accurate description in the resonant-scattering regime. It is highlighted here that the numerically calculated diffusion coefficients can be polluted for low energies due to missing resonant interaction possibilities
of the particles with the turbulence. Five reduced-rigidity regimes are established, which are separated by analytical boundaries derived in the present work.
Consequently, a proper description of cosmic-ray propagation can only be achieved by using a turbulence-level-dependent diffusion coefficient and can contribute to solving the Galactic cosmic-ray gradient problem.
\end{abstract}




\begingroup
\let\clearpage\relax
\endgroup
\newpage

\section{Introduction}
Cosmic rays and their radiative emissions are virtually ubiquitous in star-forming galaxies and active galactic nuclei \citep{Grenier2015, gaggero2015,gaggero_prl2017}.  Interpreting the observations to unveil the origins of cosmic rays, their quantitative properties, how they exchange energy and momentum with their environments, and what these properties reveal about the magnetic fields that confine them requires a thorough understanding of how they propagate. While there are a number of theoretical frameworks for propagation theory (see
\citet{2015PhPl...22i1502S,2017PhPl...24e5402Z}
for reviews), due to the prevalence of turbulence in astrophysical magnetic fields, understanding
spatial transport in the presence of magnetic turbulence is a key part of all of them. In general, this requires a statistical description, usually in terms of a diffusion tensor.

The trajectories of cosmic rays through turbulent magnetic fields are controlled by the Lorentz force. The transport is therefore conceptually simple. The fluctuations \textbf{b} of the magnetic field due to plasma waves, however, enable scattering processes of the cosmic rays that lead to a random walk. 
Complexity therefore arises from the chaotic nature of the turbulent magnetic field through which the charged particles propagate. This necessitates a statistical description of transport.
The evolution of the cosmic-ray distribution can often be modelled by a diffusive process in the limit of large times\footnote{The model of the magnetic field (with an isotropic Kolmogorov-type turbulence spectrum) used within this study only leads to ballistic and diffusive propagation. In general, particle propagation in turbulence offers the possibility of subdiffusive and superdiffusive transport, in which the running diffusion coefficient decreases or increases, respectively.}.

The spatial diffusion tensor appears in the diffusion equation and characterises the spatial evolution of cosmic rays in a turbulent magnetic field. 
By choosing the right reference frame, the diffusion tensor can generally be expressed in block-diagonal form, where off-diagonal elements describe curvature and gradient drifts \citep{Jokipii1977}. When, however, the turbulence is isotropic, the diffusion tensor becomes diagonal, which allows {{one}} to split the diffusion tensor into components parallel and perpendicular to the background magnetic field. 

The diagonal elements of the diffusion tensor yield, for a point source, $\kappa_{ii} = \lim\limits_{t \rightarrow \infty}{\kappa_{ii}(t)}$, where the running diffusion coefficient is defined as
\begin{equation}
    \kappa_{ii}(t) = \frac{\left\langle (x_i(t)-x_{{i}}(0))^2 \right\rangle}{{{2}}t},
\end{equation}
where the particle's position $x_i$ is specified in Cartesian coordinates.
Here, the notation $\langle...\rangle$ refers to averaging over all particles.
For isotropic fluctuations without a homogeneous background field, the diagonal elements of the diffusion tensor are identical.

The mean-square displacement can alternatively be expressed, following the Taylor-Green-Kubo (TGK) formalism \citep{Kubo1957}, as the mean square of the time-integrated
particle velocity (in one direction $v_i$):
\begin{align}
    \left \langle  (\Delta x_i)^2\right \rangle(t) = \left\langle \left( \int \limits_0^t \mathrm{d}\tau\, v_i(\tau)\right)^2\right \rangle.
\end{align}
The underlying time invariance, 
together with the transformation of coordinates, allows for the determination of the running diffusion coefficient as \citep{Shalchi2009}
\begin{align}\label{diff_TGK}
    \kappa_{ij}(t) = \int \limits_0^t \mathrm{d}\tau\, \left\langle v_i(\tau)v_j(0) \right\rangle.
\end{align}
In the limit of large times, the running diffusion coefficient converges toward a value which is defined as the diffusion coefficient.
The running momentum diffusion coefficient $D_{ij}(t)$ for this approach reads
\begin{align}\label{momentum_Diff}
    D_{ij}(t) = \int \limits_0^t \mathrm{d}\tau\, \left\langle \frac{\mathrm{d}v_i}{\mathrm{d}\tau}(\tau)\frac{\mathrm{d}v_j}{\mathrm{d}\tau}(0) \right\rangle.
\end{align}
The diagonal elements of the momentum diffusion tensor exhibit a fundamental relation with the spatial diffusion coefficient \citep{Berezinskii1990, Schlickeiser2002, Subedi2017}:
\begin{align}\label{kappa D}
\kappa_{ii} = \frac{v^4}{6D_{ii}}.
\end{align}
This relation is especially useful because the calculation of $D_{ii}$ simplifies significantly for high-energy particles, constituting an efficient way of deriving spatial diffusion coefficients for high reduced rigidities \citep{Plotnikov2011, Snodin2015}
\begin{align}
    \rho \equiv r_\mathrm{g}/l_\mathrm{c},
\end{align} 
where $l_\mathrm{c}$ denotes the correlation length (see also the definition in Sec.~\ref{sec:2.2}). The gyroradius $r_\mathrm{g} = v/\omega_\mathrm{c}$ has been introduced, defined with respect to the background magnetic field and the relativistic gyrofrequency.
It proves useful to introduce the reduced rigidity as this quantity takes into account the energy of the particles, the length scale over which the fluctuations are correlated, and the magnetic field strength. For highly relativistic particles, as considered in this study, scalings between energy and reduced rigidity apply as described in Appendix~\ref{app:B}.\\
Depending on the reduced rigidity, particles can be divided into magnetised ($\rho \ll 1$)
and non-magnetised ($\rho \gg 1$)~\citep{Istomin2018}. Magnetised particles have a small reduced rigidity, and their treatment is more complicated than that of non-magnetised particles.

Cosmic-ray diffusion is believed to be the dominant process for the transport of cosmic rays in many astrophysical environments \citep{galprop,dragon,picard}. In particular, the leaky-box model of the Milky Way predicts that the cosmic-ray energy spectrum observed at Earth is steepened by diffusion: the spectrum is composed of the ratio of the source spectrum $Q(E)\propto E^{-\alpha}$ and the diffusion coefficient $\kappa(E)\propto E^{\gamma}$
, i.e., $N(E)\propto E^{-\alpha-\gamma}$ \citep{Berezinskii1990}. These arguments are based on quasi-linear theory (QLT), where an assumed form of the turbulence spectrum 
\begin{align}
G(k) \propto k^{-\alpha},
\end{align}
with $k$ being the wavenumber,
leads to a parallel-diffusion-coefficient dependency\footnote{For non-relativistic particles, QLT predicts $\kappa_\parallel \propto E^{3/2-\alpha/2}$ (see \citet{Giacalone1999} for details).} $\kappa_\parallel \propto E^{2-\alpha}$ for highly relativistic particles as described in Sec.~\ref{sec:2.2}.


Several studies have investigated the diffusion coefficient tensor via numerical simulations in pure turbulence $B=0$ \citep{Parizot2004,  Globus2007, Fatuzzo2010, Plotnikov2011, Harari2013, Harari2015, Giacinti2017, Subedi2017} or with a non-vanishing background field $B$ for varying ratios of $b/B$ \citep{Giacalone1999, Casse2001, Parizot2004, DeMarco2007,Fatuzzo2010, Plotnikov2011,Harari2013, Harari2015,Snodin2015,  Subedi2017,  Giacinti2017}. Most of these results were interpreted in such a way that the numerically calculated diffusion coefficient dependencies were consistent with the predictions of QLT, independent of the turbulence level $b/B$. \citet{Minnie2007}, however, pointed out for the case of a composite of slab and two-dimensional (2D) fluctuations that turbulence-level dependency is expected and that the QLT result is only recovered for small turbulence levels.  In addition, recent studies (e.g., \citet{Snodin2015}) state that the range of energies considered for determining the diffusion coefficient dependencies is important. In the present study the resonant scattering range is further constrained by introducing a lower limit based on physical considerations, which is more restrictive than in previous numerical work. This improvement makes it possible to study the turbulence-dependent slopes of the diffusion coefficients and subsequently to quantify the findings from \citet{Minnie2007} for Kolmogorov-type turbulence. One of the goals of the present paper is to introduce a new propagation regime that exists below the energy range of the resonant scattering regime and above the non-resonant scattering regime.\\

This paper is organised as follows. Section~\ref{sec:2} presents theoretical diffusion-coefficient dependencies for both the weak- and the strong-turbulence limit. Section~\ref{sec:3} provides a recipe for the calculation of diffusion coefficients and introduces a physical lower limit of the resonant scattering regime. Section~\ref{sec:4} applies the physical constrains of the resonant scattering regime regarding the reduced-rigidity range and quantifies the turbulence-dependent spectral behaviour of the diffusion coefficients.

\section{Summary of Previous Results for the Spatial Diffusion Coefficient Dependencies}\label{sec:2}
\subsection{Diffusion Coefficients for Small Reduced Rigidities}\label{sec:2.2}
A common approach for the calculation of diffusion coefficients for magnetised particles in turbulence is quasi-linear theory (QLT), proposed by \citet{Jokipii1966}, and
its generalisations, see \citet{Matthaeus2003, Shalchi2009,Shalchi2009AA}.
Within QLT, the particle motion is assumed to be a superposition of the gyromotion of the particle and stochastic motion of the guiding centre along magnetic field lines. 
The motion of the particle is modelled by the unperturbed trajectory. This simplification can, however, only be justified in the limit of $b \ll B$.\\
An additional assumption is the gyroresonance condition, which implies that particles only interact resonantly with fluctuations at a fixed wavelength $l$ that is determined via
\begin{align}\label{resonantScattering_0}
|\mu| = \frac{l}{2\pi r_{\mathrm{g}}},
\end{align}
where $\mu = \cos\Theta_0$ is defined as the cosine of the pitch angle $\Theta_0$. The pitch angle is defined as the angle between the particle velocity and the background magnetic field.\\\\
There is, however, the well-known problem that interactions with $\mu = 0$ are prohibited due to the resonance condition stated above \citep{Tautz2008}. 
Nonlinear transport theories have been proposed to solve this problem by replacing the sharp resonance between waves and particles with a resonance-broadening function \citep{Voelk1973,Jones1973,Goldstein1976,Shalchi2004, Yan2008, Mertsch2019} or by taking into account fluctuations in the electric field \citep{Schlickeiser1989}.\\\\
Despite these problems, agreement of numerical simulations with the dependencies of the diffusion coefficient derived within the QLT formulation was found in several studies. 
In this Section, the expected dependencies in the formalism of QLT are presented.\\\\
In QLT, $\kappa_\parallel$ is inversely proportional to the scattering rate $\nu_\mathrm{s}$ \citep{Berezinskii1990, Schlickeiser2002}, as per
\begin{align}\label{kappa_diff}
\kappa_\parallel = \frac{v^2}{4} \int \limits_0^1 \mathrm{d}\mu \frac{1-\mu^2}{\nu_\mathrm{s}}\propto v^2\nu_\parallel^{-1}.
\end{align}
The scattering rate can be approximated within the formulation of QLT as \citep{Kulsrud1969, Berezinskii1990, Zweibel2013}
\begin{equation}
  \nu_{\parallel}\approx 2\pi^2\left|\omega_B \right|\,\frac{k_{\rm res}\,\varepsilon(k_\mathrm{res})}{B^{2}}\,,
  \label{scattering_rate:equ}
\end{equation}
where $\omega_B$ denotes the synchrotron frequency, which is proportional to the resonant wavenumber $k_\mathrm{res}$. The wave energy at wavenumber $k_\mathrm{res}$ is $k_{\rm res}\,\varepsilon(k_\mathrm{res})$ \citep{Zweibel2013}.
Isotropic turbulence with a Kolmogorov spectrum in the inertial range is considered, assuming that energy is injected at $l_\mathrm{max}$ and dissipated at $l_\mathrm{min}$ after a cascade from large to small wavelengths without energy loss.
Under these assumptions, the turbulent spectrum {{$G(k)$}} follows a power law with the spectral index $\alpha = 5/3$ in one dimension,
{{
\begin{align}
G(k) = \frac{b^2}{8\pi}\, k^{-\alpha}\, \frac{(\alpha-1)\,k_{\rm min}^{\alpha-1}}{1-(k_{\rm max}/k_{\rm min})^{\alpha-1}}\,,
\end{align}}}
which is expected to be applicable for various astrophysical environments such as jets \citep{Casse2001}.
For a one-dimensional Kolmogorov spectrum with $\alpha = 5/3$, in the limit $l_\mathrm{min} \ll l_\mathrm{max}$ the correlation length\footnote{This definition differs from another definition of the correlation length that is frequently used in the literature \citep{Monin1975}:\\
        $l_\mathrm{c} =   \int_0^\infty \mathrm{d}rR(r)/R(0)=\pi/2   \int_0^\infty \mathrm{d}k\,G(k)k^{-1}/ \int_0^\infty \mathrm{d}k\,G(k)\approx l_\mathrm{max}/10$ in the limit $l_\mathrm{min} \ll l_\mathrm{max}$. All subsequent given correlation lengths are calculated according to $l_\mathrm{c}=l_\mathrm{max}/5$ or converted to it if they are quoted from other papers that have chosen a different definition (see for example overview Tab.~\ref{paper}).} approximately yields \citep{Harari2013}
\begin{align}\label{corr_l_max}
l_{\mathrm{c}} = \frac{l_{\rm max}}{2}\,\frac{\alpha-1}{\alpha}\,\frac{1-(l_{\rm min}/l_{\rm max})^{\alpha}}{1-(l_{\rm min}/l_{\rm max})^{\alpha-1}}\approx \frac{l_\mathrm{max}}{5}.
\end{align}\\\\
These expressions can be inserted back into Eq.~(\ref{kappa_diff}), yielding
\begin{align}\label{diff_regime_dep}
\kappa_\parallel \approx \frac{2\pi v}{k_\mathrm{c}} \left(\frac{k_\mathrm{res}}{k_{\mathrm{c}}}\right)^{\alpha -2} \frac{B^2}{b^2}  = vl_\mathrm{c}\,\rho^{2-\alpha}\frac{B^2}{b^2},
\end{align}
{{where the correlation wavenumber $k_\mathrm{c}$ is connected to the correlation length via $l_\mathrm{c}=2\pi/k_{\rm c}$. This equation reproduces the parallel diffusion coefficient of QLT in the limit of $b \ll B$.}} \\\\
In the limit $b \gg B$, the particle orbits are not close to the unperturbed trajectories anymore. 
Recent developments have improved the understanding of the parallel-diffusion-coefficient dependencies in this particular limit \citep{Casse2001, Shalchi2009, Harari2013, Snodin2015, Subedi2017}. In the strong-scattering limit within nonlinear diffusion theory \citep{Shalchi2009}, the modified Bohm limit 
yields \citep{Srinivasan2014, Hussein2014}
\begin{align}\label{eq:bohm}
    \lambda_\parallel = \frac{r_\mathrm{g}}{2} \frac{B}{b}.
\end{align}
Like the original Bohm limit, the modified Bohm limit describes the proportionality between the mean-free path $\lambda_\parallel = 3\kappa_\parallel/c$ and the gyroradius; however, it corrects for the influence of the turbulence. The resulting parallel diffusion coefficient is independent of the mean magnetic field and reads
\begin{align}
    \kappa_\parallel = \frac{E}{6qb},
\end{align}
where $q$ denotes the charge of the particle and $E$ its energy.
Consequently, the parallel-diffusion-coefficient dependencies can be derived in both the quasi-linear limit and the Bohm limit for relativistic particles with speed $v=c$ and expressed as functions of the reduced rigidity and the ratio $b/B$,
\begin{align}\label{large_b}
\kappa_\parallel = \begin{cases}
cl_\mathrm{c}\,\rho^{2-\alpha} B^2/b^2 &    $for $  b \ll B\\
c\, l_\mathrm{c}\rho  B/(6b) &    $for $ b \gg B
\end{cases}.
\end{align}
The weakly nonlinear theory in the small-gyroradius limit (see \citet{Shalchi2004} for details) was developed to describe the diffusion coefficient dependencies between these $b/B$ limits.

While some studies \citep{Casse2001, Fatuzzo2010, Giacinti2017} have found agreement of their simulation results with the predictions of QLT even for strong turbulence, i.e., no agreement with the predictions of Bohm-like diffusion according to Eq.~(\ref{eq:bohm}), \citet{Snodin2015} have found Bohm-like diffusion of particles for a large energy range. In the latter work, linear energy scaling of the parallel diffusion coefficients was observed for particles with low reduced rigidities. An overview of the results of previous papers can be found in Tab.~\ref{paper}. In Sec.~\ref{sec:4}, they are contrasted with the present results, whose energy behaviour also agrees with the Bohm-like diffusion prediction for strong turbulence levels.

\subsection{Diffusion Coefficients for Large Reduced Rigidities}\label{Large_Rho}
For particles with gyroradii that substantially exceed the correlation length of the turbulence, their direction is expected to change only slightly over a correlation length. In the limit of $\rho \gg 1$, the relative magnitude of this change can be approximated by $1/\rho$. In \citet{Plotnikov2011}, diffusion-coefficient dependencies are derived using a Markovian description of the trajectories. As an alternative, the dependencies of the diffusion coefficient for turbulence without a background field are derived in \citet{Subedi2017} using the connection between velocity-space diffusion and spatial diffusion. The change of momentum is described by the Lorentz force
\begin{align}\label{lorentz}
\frac{\mathrm d \textbf{{v}}}{\mathrm{d} t} = \frac{q}{m\gamma(v)}(\textbf{{v}} \times (\textbf{{b}}+\textbf{{B}})),
\end{align}
with mass $m$, speed $v$, Lorentz factor $\gamma(v)$, and charge $q$ of the particle. Inserting the Lorentz force into the momentum diffusion coefficient in Eq.~(\ref{momentum_Diff}) within the TGK formalism for large times results in the expression
\begin{equation}\label{Diff_Momentum}
\begin{split}
    D_{ij} = \left(\frac{q}{m \gamma(v)}\right)^2 &\lim\limits_{t \rightarrow \infty} \int \limits_0^t\mathrm{d}\tau\,\epsilon_{i\alpha\beta} \epsilon_{j\gamma\eta} \\ \cdot
    &\left\langle v_\alpha(0)v_\gamma(\tau)\bigl(b+B\bigr)_\beta(0)\bigl(b+B\bigr)_\eta[x(\tau)]\right\rangle.
    \end{split}
\end{equation}
Following the argumentation of \citet{Matthaeus2003, Subedi2017}, the local particle velocity is uncorrelated from the local magnetic field vector only when there is an isotropic particle distribution and when the turbulence is statistically homogeneous. The arguments are based on the Corrsin approximation (see \citet{Tautz2010}), which is essentially a random-phase approximation.
These conditions are fulfilled for statistically isotropic turbulence in three dimensions without a background field. The velocity correlation yields $\left\langle v_\alpha (0)v_\gamma(t)\right\rangle = v_\alpha v_\gamma$. The remaining integral can be interpreted as the squared magnitude of the fluctuations divided by the speed of light times the correlation length $l_\mathrm{c}$, which is defined for particles with large reduced rigidities as
\begin{align}\label{correlation_Length}
    l_\mathrm{c} = \frac{c}{b^2}\int \limits_0^\infty \mathrm{d}t \left\langle b_i(x( t))b_i(x(0))\right\rangle,
\end{align}
where $b_i$ are the turbulent fluctuations in the Cartesian coordinate system.
Using Eq.~(\ref{kappa D}), the diagonal elements of the diffusion tensor are determined to be
\begin{align}
    \kappa_{ii} = \frac{1}{2}
    \rho^2 c \,l_{\mathrm{c}},
\end{align}
where $r_g \propto 1/b$ is utilized due to the missing background field.
For the case of an additional background magnetic field, particles are only isotropically distributed within the plane perpendicular to the background magnetic field vector.
The parallel momentum diffusion coefficient in Eq.~(\ref{Diff_Momentum}) only takes into account the perpendicular velocity components, while the perpendicular momentum diffusion coefficient is based on the parallel and the perpendicular particle velocity distributions. Consequently, the abovementioned condition is only fulfilled for the parallel component of the momentum diffusion coefficient, which yields
\begin{equation}
\begin{split}
    D_\parallel = \left(\frac{q}{m \gamma(v)}\right)^2&\lim\limits_{t \rightarrow \infty} \int \limits_0^t\mathrm{d}\tau\,\epsilon_{3\alpha\beta} \epsilon_{3\gamma\eta}\\ &\cdot \left\langle v_\alpha(0)v_\gamma(\tau)\right\rangle\left\langle b_\beta(0)b_\eta[x(\tau)]\right\rangle.
    \end{split}
\end{equation}
The velocity correlation is proportional to the perpendicular velocity $v_\perp$ of the particles as long as their trajectories can be treated as unperturbed. Combining these assumptions and using the relation between the spatial and momentum diffusion coefficient results in
\begin{align}
\kappa_\parallel \propto \left( \frac{B}{b} \right)^2  \rho^2 c \,l_{\mathrm{c}},
\end{align}
which is in agreement with the result of derivation using a Markovian description \citep{Plotnikov2011}.

\section{Considerations in Numerical Simulations of Diffusion}\label{sec:3}

The main challenge in investigating the diffusion coefficient's parametric dependencies numerically arises from the necessity for simulating a large range of particle energies.
It is difficult to preserve the numerical convergence of the simulated diffusion coefficients over the entire range of particle energies, given that the particle energy determines the range of plasma wavelengths with which the particles can resonantly interact, i.e., $l = |\mu|2\pi r_{\mathrm{g}}$.
 
As a consequence, the range of wavelengths $l$ of the fluctuations $b$ has to extend well below the gyroradius of the lowest-energy particle and up to the gyroradius of the highest-energy particle.
In order to cope with this large range of scales, simulations generally employ a synthetic random magnetic field, either composed of a superposition of static plane waves in Fourier representation \citep{Snodin2015, Giacinti2017} or specified on a discrete mesh \citep{Giacinti2012, DeMarco2007, Giacalone1999}. Both methods correspond to different ways of specifying the same model with different shortcomings, especially for low reduced rigidities. Whereas with the second method the resolution of the magnetic field is limited by the available memory, the required computing time scales with the number of modes taken into account for the superposition in the first method.\\\\
The relation between the simulation parameters and the resulting diffusion coefficient is multilayered and highly entangled. Subtle details of the magnetic field structure, such as the magnetic mode density \citep{Snodin2015}, together with the range of wavenumbers involved, influence the simulated diffusion coefficient, as will be demonstrated here.

\subsection{Test-Particle Simulation Setup}\label{setup}
Test-particle simulations were performed within the CRPropa framework\footnote{{{The specific version used for the simulations is CRPropa 3.1-f6f818d36a64.}}}, which is a publicly available tool for simulations of cosmic-ray transport and its secondaries \citep{AlvesBatista2016}. The numerical framework employed here restricts our analyses to the highly relativistic limit, but many conclusions apply to the general case, as well. Specifically, we have replaced $v$ by $c$ in relating the diffusion coefficient $\kappa$ to the pitch angle scattering coefficient $\nu$ and in defining the gyroradius $r_\mathrm{g}$. Note that the Lorentz force equation depends on particle charge $q$, rest mass $m$, and Lorentz factor $\gamma$ in the combination $q/(m\gamma)$, so our results can easily be generalized in this respect as well. \\\\
Our simulation framework is based on a modular architecture and provides various interaction, observer, deflection,
and boundary modules. The Boris push method \citep{Qin2013,Winkel2015} is used for propagating mono-energetic charged particles within a magnetic field.
This method resolves the velocity dependence in the equations of motion,
stated by the Lorentz force.
Due to its fast computation and long-term
precision, it is widely used for advancing a charged particle within a magnetic field \citep{Qin2013, Winkel2015}. In Sec.~\ref{Numerical_Influence}, the convergence properties of this method are investigated.\\\\ 
The diffusion time of relativistic charged particles interacting with hydromagnetic waves is much shorter than the time scale of 
acceleration effects \citep{Fatuzzo2010}.
As a consequence, electric fields are neglected, and magnetic fields are set to be stationary. The regular field $\textbf{{B}}$ is chosen to be aligned with the $x_3$-axis, i.e., {{$\textbf{{B}} = B \, \textbf{{e}}_{3}$, with $B = |\textbf{B}| = 1\,\mu\mathrm{G}$}}. The synthetic random magnetic field is specified on a discrete mesh, and the complex turbulent magnetic field vectors $\textbf{{b}}(\textbf{{k}})$ are first defined on a regular grid in three-dimensional
wavenumber space as 
{{
\begin{align}\label{b_tubr}
\begin{split}
\textbf{{b}}(\textbf{{k}}) = &\chi(k) G(k)^{1/2}\\ &\cdot [\textbf{{e}}_1(\textbf{{k}}) \cos(\Theta(\textbf{{k}}))+\textbf{{e}}_2(\textbf{{k}}) \sin(\Theta(\textbf{{k}}))] \exp(i\Phi(\textbf{{k}})),
\end{split}
\end{align}
}}
where $\textbf{{e}}_1(\textbf{{k}})$ and $\textbf{{e}}_2(\textbf{{k}})$ are orthonormal vectors confined to the plane perpendicular to the wavevector $\textbf{{k}}$. The {{ orientation of the vectors $\textbf{{e}}_1(\textbf{{k}})$ and $\textbf{{e}}_2(\textbf{{k}})$ is defined by the random phase $\Theta(\textbf{{k}})$, and the random variable $\Phi(\textbf{{k}})$ determines the real and imaginary proportion.}} In addition, $\chi(k)$ is introduced to guarantee the mean of $\textbf{{b}}(\textbf{{k}})$ to be zero\footnote{While there is growing evidence that MHD turbulence is anisotropic (see for example \citet{Sridhar1994}), we defer consideration of this hypothesis to future work.}.
The normal base $\textbf{{k}}_\mathrm{n}/k_\mathrm{n}$, $\textbf{{e}}_1$, $\textbf{{e}}_2$ ensures that ${\nabla} \cdot \textbf{{b}} = 0$. The turbulent magnetic field on a regular, three-dimensional Cartesian grid is generated using the inverse Fourier transform of Eq.~(\ref{b_tubr}){{ and is afterwards re-adjusted to the specified root-mean-square value for the turbulent component}}.\\\\
Discrete storage of the turbulent magnetic field $\textbf{b}$ on a regular grid with $N_\mathrm{grid}^3$ grid points and isotropic spacing $s_{\mathrm{spacing}}$ constrains the possible plasma waves that can fit into the box, subject to the conditions
\begin{align}
l_\mathrm{min}\geq 2 \,{s_{\mathrm{spacing}}},\label{l_cond}\\
l_\mathrm{max} \leq N\mathrm{_{grid}} \, s_{\mathrm{spacing}}/2 \leq N\mathrm{_{grid}}l_\mathrm{min}/4,\label{l_cond1}
\end{align}
where $l_\mathrm{min}$ is defined as the smallest numerically resolved wavelength, and $l_\mathrm{max}$ represents the largest wavelength of the plasma waves that are allowed by the simulation. However, in order to ensure isotropic turbulence even at large wavelengths, averaging over many simulations with different realisations using the same parameters is necessary.
The magnetic field at an arbitrary trajectory position between grid points is obtained by linear interpolation. Numerical interpolation effects are briefly discussed in Sec.~\ref{Numerical_Influence}. With those constraints on the possible range of plasma wavelengths, the energy spectrum $G(k)$ for wavenumber $k$ = $2\pi/l$ is given by
\begin{align}
G(k) \propto 
\begin{cases} 
     0  & \text{if } k < k_\mathrm{min},\\
   \left(k/k_{\mathrm{min}}\right)^{-\alpha}    & \text{if } k_\mathrm{min} \leq k \leq k_\mathrm{max} \\
   0   & \text{if } k_\mathrm{max} < k
  \end{cases},\label{eq:spectrum}
\end{align}
where $\alpha$ is the spectral index. \\\\
The gyroradius $r_\mathrm{g}$ in numerical simulations is defined as 
\begin{align}\label{gyroradius_sim}
r_{\mathrm{g}} = \frac{E} {cqB },
\end{align}
in accordance with \cite{Candia2004,DeMarco2007}.

\subsection{Temporal Convergence of the Running Diffusion Coefficient}

The running diffusion coefficient can be calculated with different methods as summarised in Tab.~\ref{tab:methods}. For diffusive transport, the running diffusion coefficient converges to the diffusion coefficient for $t \rightarrow \infty$. The diffusive regime starts as soon as the particles are completely decorrelated from their initial condition, caused by chaotic fluctuations. In the following, the Second Moment method is applied for the calculation of diffusion coefficients.
\begin{table}
    \centering
	\caption{Methods for the numerical evaluation of the diffusion coefficient.}\label{tab:methods}
	\begin{tabular}{|ccc}
		\hline 
		Method & Calculation & Eq.\tabularnewline
		\hline 
		TGK Formalism &  $\kappa_{ii}  = \sum_{n = 0}^{t/(\Delta t)} \left\langle v_i(n\cdot\Delta t)\Delta x_i(0) \right\rangle$ & \refstepcounter{equation}(\theequation)\tabularnewline
		Diffusion Equation & $\kappa_{ii}=\sigma/(2t)$ & \refstepcounter{equation}(\theequation) \tabularnewline
		Second Moment &  $\kappa_{ii} = \lim\limits_{t \rightarrow \infty} \langle(\Delta x)^2\rangle/(2t)$ &\refstepcounter{equation}(\theequation) \tabularnewline
		\hline 
	\end{tabular}
\end{table}

\begin{figure}
\includegraphics[width=\columnwidth]{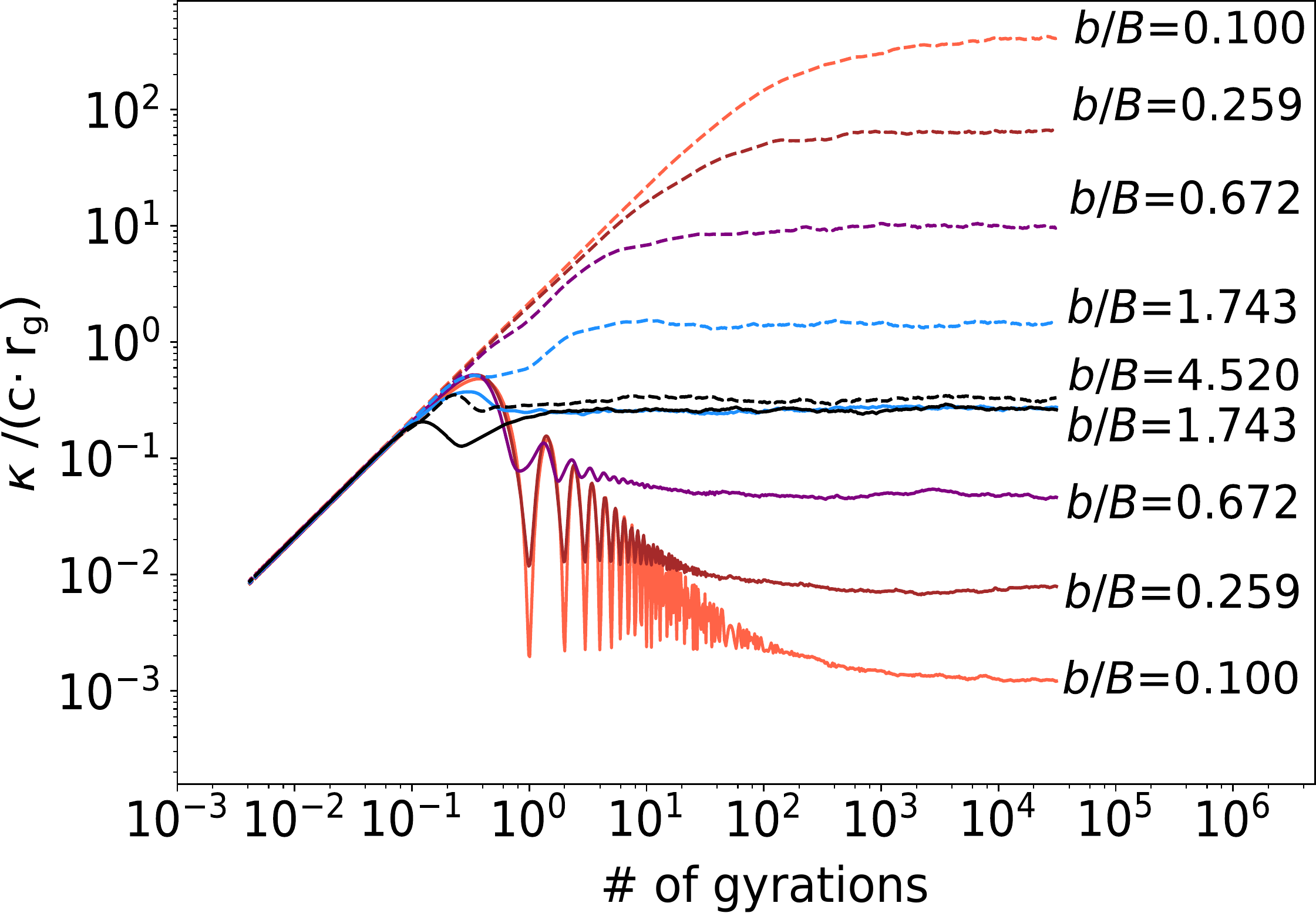}\caption{Normalised diffusion coefficients (parallel are dashed and perpendicular are solid) as functions of the number of gyrations at a proton energy of $10^5$ TeV. The running diffusion coefficient converges to the final diffusion coefficient for large times. Once this plateau is reached, it is a diffusive process. Prior to this, the gyration motion of the individual particles dominates the transport due to the background field and leads to temporally proportional and inversely proportional behaviour of the parallel and the perpendicular diffusion coefficient, respectively. 
Different turbulence levels demonstrate the influence of this quantity on the time scales and the diffusive transport. Simulated with $l_\mathrm{min} = 1.7~\mathrm{pc}$,
		$l_\mathrm{max} = 82.5\,\mathrm{pc}$,
		$s_{\mathrm{spacing}} = 0.17\,\mathrm{pc}$, $N_{\mathrm{grid}} = 1024$, $B = 1\, \mu\mathrm{G}$, and one magnetic field realisation each.}\label{fig:10000}
\end{figure}

\begin{figure}
\includegraphics[width=\columnwidth]{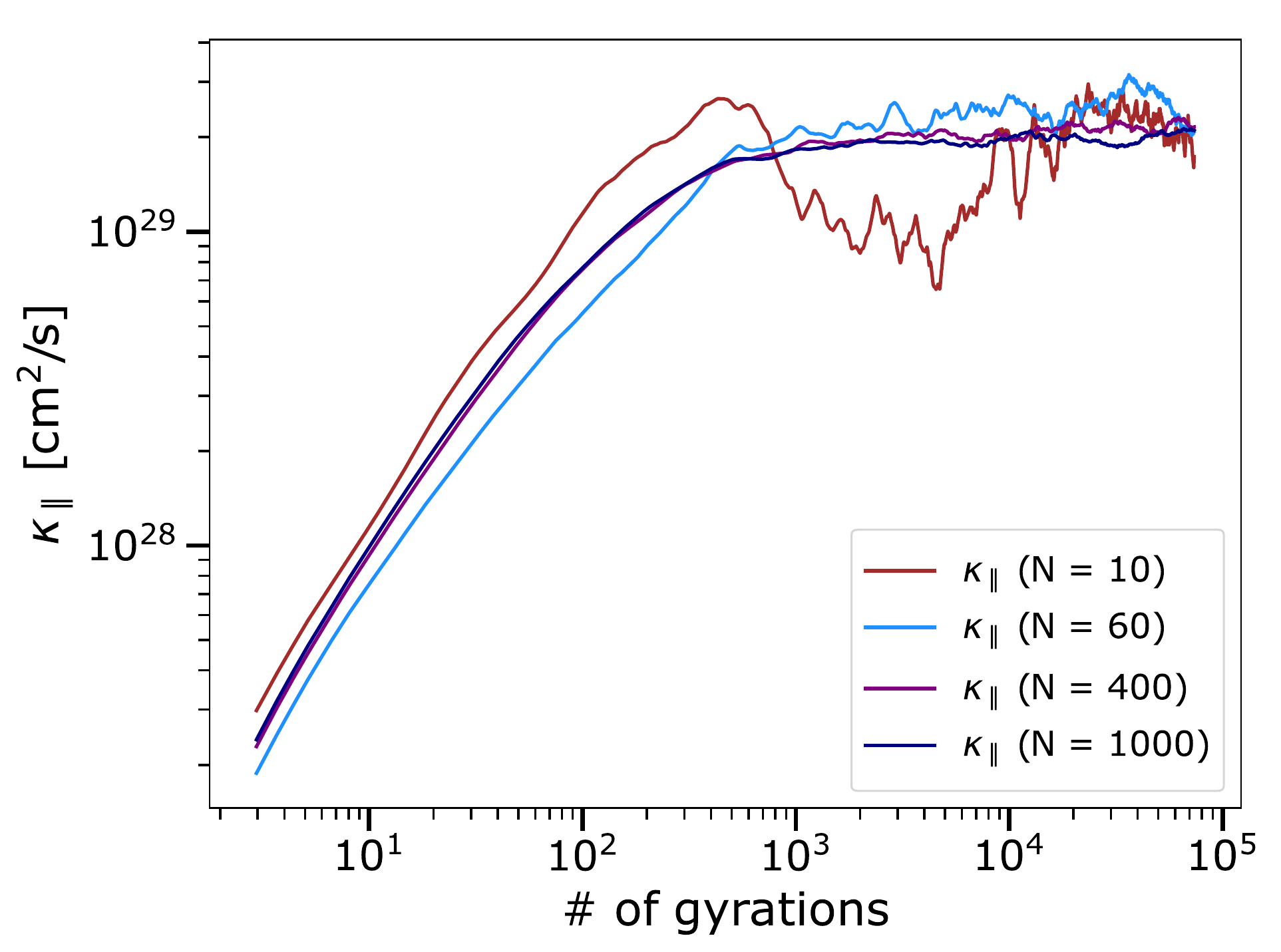}\caption{Running diffusion coefficient for simulations with different numbers of grid points and particles with $E=10^5~\mathrm{TeV}$. Running diffusion coefficients that are computed with more than 100 particles converge to a stable plateau. Simulated with $l_\mathrm{min} = 1.7~\mathrm{pc}$,
		$l_\mathrm{max} = 82.5\,\mathrm{pc}$,
		$s_{\mathrm{spacing}} = 0.17\,\mathrm{pc}$, $N_{\mathrm{grid}} = 1024$, $b = 0.1\,\mu\mathrm{G}$, $B = 1\, \mu\mathrm{G}$, and one magnetic field realisation.}\label{fig:numberpa1}
\end{figure}

Figure~\ref{fig:10000} presents both components of the normalised diffusion coefficients as functions of the number of gyrations for different turbulence levels $b/B$. 
The plateau of the running diffusion coefficient can be identified with the diffusion coefficient and does not appear before the chaotic character of the trajectories dominates the gyromotion due to the background field. 
Consequently, the running diffusion coefficient can be classified into two temporal regimes:
\begin{enumerate}
	\item \textbf{Weakly-Perturbed-Propagation Regime}: 
	{{At early times $t$, }} the parallel running diffusion coefficient yields 
	\begin{align}
	\kappa_{\parallel}(t) = \frac{\langle(\Delta x_3)^2\rangle}{2t} \propto t,
	\end{align}
	resulting in the linear increase of the running diffusion coefficient seen for the dashed lines in Fig.~\ref{fig:10000}.
	The turbulent magnetic field, however, causes a slight displacement of the particle after each gyration, such that the running perpendicular diffusion coefficient is not vanishing at its local minima after each gyration, even for high temporal resolution. This wiggling effect is observed in Fig.~\ref{fig:10000} for the solid lines for the first gyrations until the plateau is reached. 
	{{While in Fig.~\ref{fig:10000}, one may conclude that the perpendicular running diffusion coefficient is subdiffusive, this effect is actually due to the gyration motion. Since the transport is initially dominated by the background field, the perpendicular spatial expansion remains constant, so that the resulting diffusion coefficient exhibits the characteristic inversely proportional decrease in time 
    \begin{align}
        \kappa_{\perp}(t) \approx \frac{\langle(\Delta x_{1})^2+(\Delta x_{2})^2\rangle}{2t}  \propto \frac{1}{t}.
    \end{align}
	}} 
	\item \textbf{Diffusive-Propagation Regime}: For large times, the trajectories are mainly influenced by the turbulent magnetic field and therefore best characterised by chaotic movement. In this limit, the running diffusion coefficients are constant for both the parallel as well as the perpendicular component. The distance travelled before diffusion starts is approximately one mean-free path:
	\begin{align}\label{start_plateau}
	\lambda_\parallel = \frac{3\kappa_\parallel}{c},
	\end{align}
	which refers to the distance between two scatterings off magnetic perturbations.
	After a distance $\lambda_\parallel$, the direction of
	the particle is statistically decorrelated from the
	initial direction. 
\end{enumerate}
An insufficient number of particles may either prevent the running diffusion coefficient from reaching a plateau or add artificial quasi-chaotic movement. In addition, the analysis of too few particles may introduce subdiffusive or superdiffusive regimes, instead of the appropriate diffusive behaviour. In Fig.~\ref{fig:numberpa1}, the running parallel diffusion coefficients are presented as functions of the number of completed gyrations, for different numbers of particles. The quasi-chaotic movement of the running diffusion coefficient for few particles in the diffusive-propagation regime is due to an insufficient number of particles. An increased number of particles does not only stabilise the plateau but also helps to find the transition between an increasing running diffusion coefficient and its plateau.

\subsection{Convergence of the Diffusion Coefficient}
As demonstrated above, in order to reach a stable plateau of running diffusion coefficients, the trajectory length must be sufficiently long and the number of particles sufficiently high. However, finding a plateau for certain simulation parameters, such as the box size and the step length, does not guarantee that the plateau is numerically converged. To ensure that the time-converged running diffusion corresponds to the numerical converged final diffusion coefficient, further conditions are required to be fulfilled. The diffusion coefficient is only expected to recover the physical result if its value is numerically converged, which means that its value remains essentially unchanged upon increase of numerical resolution or particle number.\\
Before these numerical parameters are discussed in Sec.~\ref{Numerical_Influence}, the different diffusion ranges are discussed in the following, since these can be directly influenced by some simulation parameters. The calculated diffusion coefficients are only physically meaningful in the context of the following regimes.

\subsubsection{Resonant scattering}\label{sec:3.1.1}
Particles with pitch angle $\mu$ and gyroradius $r_\mathrm{g}$ interact according to the resonance criterion with fluctuations of size {{$l=2\pi |\mu| r_\mathrm{g}$}}, with a resultant change in pitch angle $\delta\mu$ of order $\delta \mu \approx b/B$ \citep{Kulsrud1969}. 
Treating continuous scattering as a random walk requires a sufficient density of waves such that a particle can jump from one wave to another. Any particle
will run out of resonant waves when its {{$|\mu|$}} is small enough that the resonant wavenumber is greater than $k_\mathrm{max}$. Then, mirroring can take over if $b/B$ is large enough.
The fluctuations which form the mirror will generally be of longer wavelength, as they have larger amplitude and also maintain the adiabatic invariance of the magnetic moment.

\begin{figure}
	\includegraphics[width=\columnwidth]{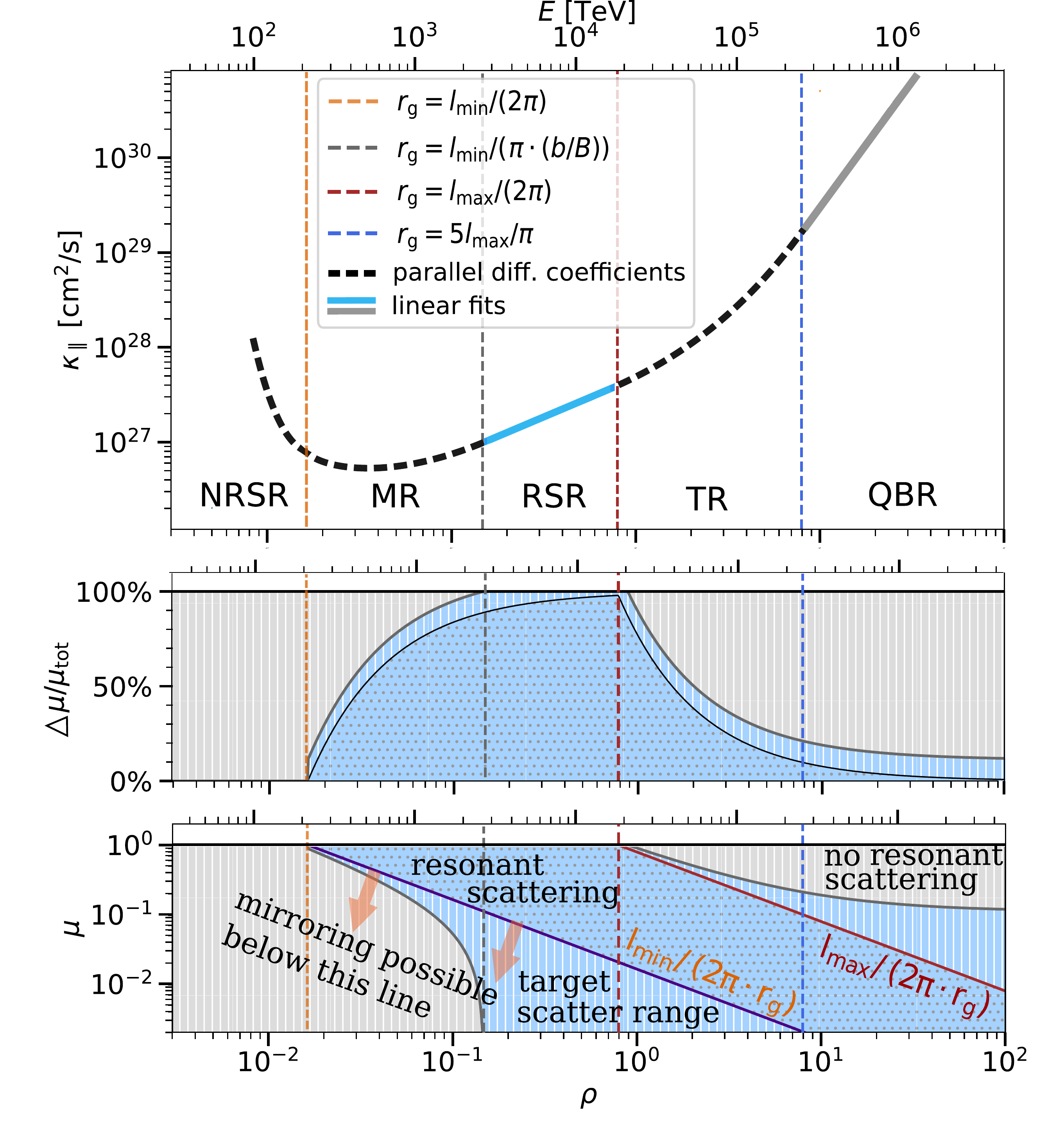}\caption{\textit{Upper panel}: Schematic plot of parallel diffusion coefficients as functions of the reduced rigidity. The different regimes result from the portion of the pitch angle with which particles of a certain $\rho$ can scatter resonantly at the plasma waves. \textit{Middle panel}: The graph illustrates the percentage of the range of $\Delta \mu$ that is accessible for resonantly interacting particles. $\mu_\mathrm{tot}$ denotes the range of all possible values $-1 \leq \mu \leq 1$. 
	\textit{Lower panel}: Illustration of the gyroresonance condition, which is fulfilled only in the blue dotted area. 
	The blue dashed area indicates parameter combinations that can be reached via scattering by particles within the blue dotted area through scattering with $\Delta \mu \approx b/B$. {{The parameters are $l_\mathrm{min} = 1.7\,$pc, $l_\mathrm{max} = 82.5\,$pc, and $b/B=0.1.$}}}
	\label{fig:influence_scatter}
\end{figure}

\begin{figure}
	\includegraphics[width=\columnwidth]{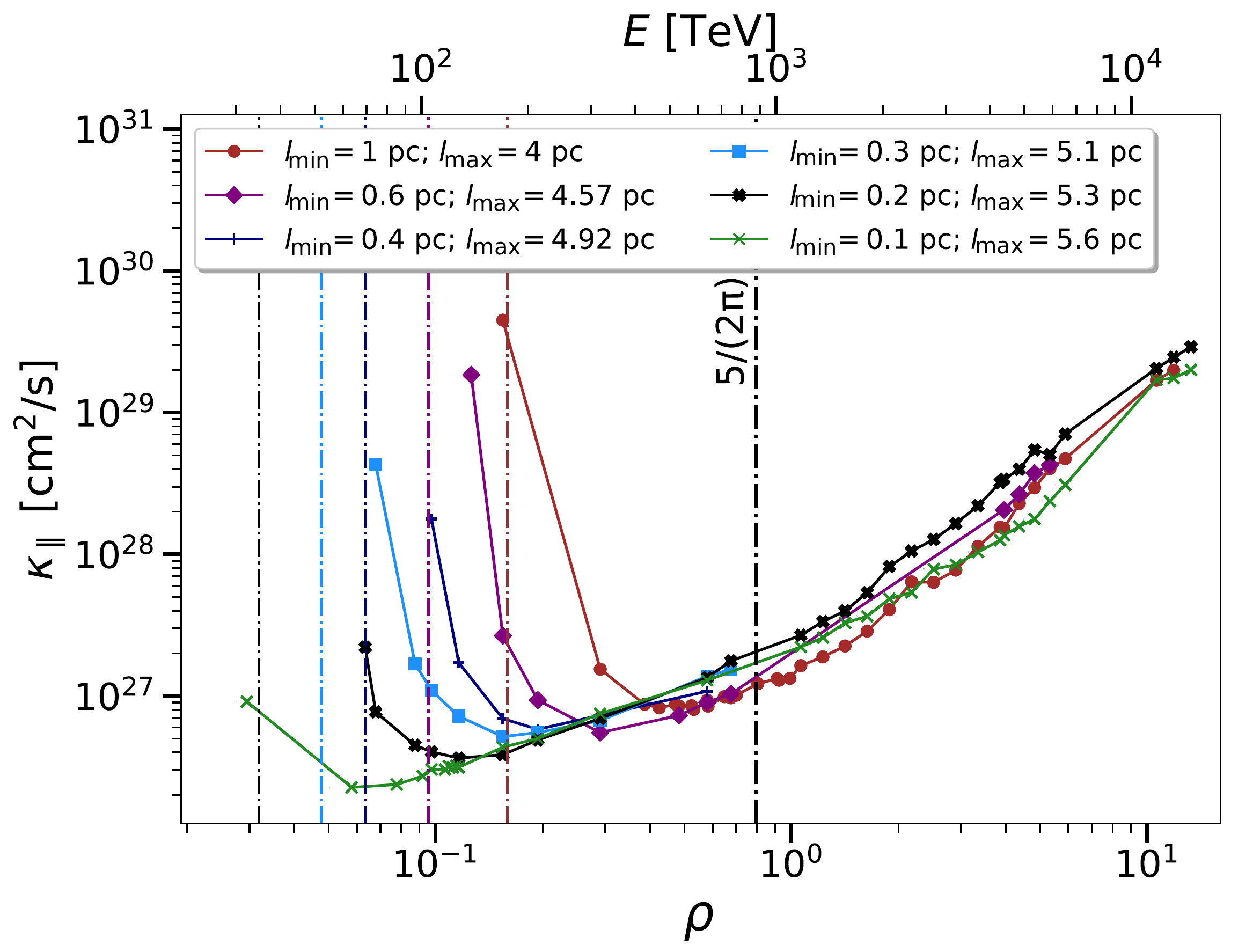}\caption{Diffusion coefficients as functions of the reduced rigidity for different ranges of fluctuation scales $l$. The black dash-dotted line represents the upper boundary of the RSR (see Tab.~\ref{tab:theoretical predictions}), while the coloured dash-dotted lines indicate the boundaries between NRSR and MR for a given set of scales. Simulated with a turbulent field together with a background field. Simulated with $s_\mathrm{spacing} = 0.17$ pc, {{$N_\mathrm{grid} = 1024$}}, $b=0.1~\mathrm{\mu G}$, $B=1~\mathrm{\mu G}$, $l_\mathrm{c} = 1.2$ pc.}\label{fig:l_min}%
\end{figure}

\begin{figure}
	\includegraphics[width=\columnwidth]{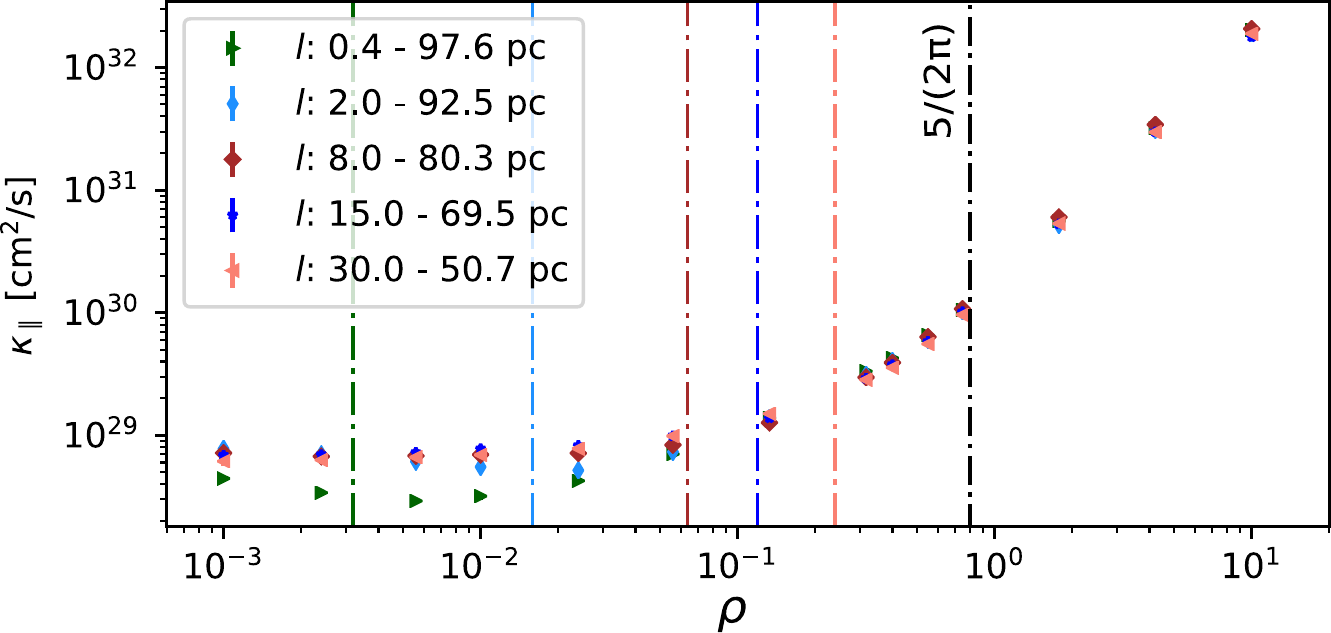} \caption{Diffusion coefficients as functions of the reduced rigidity for different ranges of fluctuation scales $l$. The black dash-dotted line represents the upper boundary of the RSR, while the coloured lines indicate the boundaries between NRSR and MR. Simulated with a purely turbulent field without {{a}} background field. The MR and RSR coincide for the case of a vanishing background magnetic field. The energy independent FLRW dominates the transport of low energetic particles. The deviation from this behaviour for the green triangles is caused by the interpolation routine, which generates a artificial background magnetic field between grid points. Simulated with $s_\mathrm{spacing} = 0.17$ pc, {{$N_\mathrm{grid} = 1024$}}, $b=0.1~\mathrm{\mu G}$, $B=0$, $l_\mathrm{c} = 20$ pc. }\label{fig:b_l_min}%
\end{figure}

The establishment of different reduced-rigidity regimes is based on physical considerations and requires the detailed investigation of the possible resonant scattering interactions of the particles given a certain range of fluctuation wavelengths. The parallel diffusion coefficient within the resonant scattering regime (RSR) is presented in the upper panel of Fig.~\ref{fig:influence_scatter} as a solid blue line. This is the only regime where QLT predictions are applicable. A slope of 1/3 {{for}} this blue line is expected in the limit $b\ll B$. The two main limiting assumptions of QLT, namely the need for weak turbulence levels $b\ll B$ and the gyroresonance condition, lead to strong limitations of the parameters for which QLT predictions are valid. As illustrated in the upper panel of this figure, there exist further regimes. The first systematic investigation of all reduced-rigidity regimes is presented in the following. \\\\
Figure \ref{fig:influence_scatter} shows, in the lower two panels, the parameter combinations of $\mu$ and $\rho$ for a fixed $l_\mathrm{min}$ and $l_\mathrm{max}$ for which resonant scattering is possible. The grey area indicates the parameter combinations that prohibit resonant scattering. The middle panel of Fig.~\ref{fig:influence_scatter} presents the resonant scattering range as a percentage of the total range $-1 \leq \mu \leq 1$. 
These considerations contribute to answering the following fundamental question: what is the influence of these (physical or numerical) fluctuation boundaries on the propagation of cosmic rays, and in particular on the diffusion coefficient? Based on the reduced rigidity of the particle, it can be classified as falling into one of the following regimes:
\begin{enumerate}
    \item \textbf{Non-Resonant-Scattering Regime (NRSR)}: For $\rho \lesssim l_\mathrm{min}/(2\pi\,l_\mathrm{c})$, the gyroresonance criterion reveals that particles cannot scatter resonantly, independently of $\mu$. Thus, the NRSR is defined as the range in $\rho$ for which resonant scattering is prohibited. As scattering is prohibited for the complete pitch-angle spectrum, mirroring occurs instead \citep{Cesarsky1973, felice2001, lange2013, seta2018}. 
    Figures~\ref{fig:l_min} ($B>0$) and \ref{fig:b_l_min} ($B=0$) present simulated diffusion coefficients as functions of reduced rigidity. The different fluctuation ranges demonstrate the dependence of the upper boundary of the NRSR on $l_\mathrm{min}$, indicated by the vertical dashed-dotted colored lines.
    
    \textit{For a weak turbulence level:}
    Without resonant scattering, particles follow magnetic field lines that are dominated by the strong background field. Particles can only reverse direction when encountering magnetic traps formed by the field lines. The transport is determined by the field-line geometry. The study of the influence of traps on the diffusion coefficient will be deferred to future work.
    
    \textit{For a strong turbulence level:} 
    The diffusion coefficients remain approximately constant in Fig.~\ref{fig:b_l_min} in the NRSR for strong turbulence levels due to the energy-independent field-line random walk (FLRW) that dominates without resonant scatterings. 
    \item \textbf{Mirroring Regime (MR)}: At values $l_\mathrm{min}/(2\pi l_\mathrm{c}) \lesssim \rho \lesssim l_\mathrm{min}/(\pi l_\mathrm{c} (b/B))$, the range of pitch angles that can scatter resonantly decreases towards lower reduced rigidities until the NRSR is reached. Similarly, as described in the NRSR, the behaviour for weak and strong turbulence levels is different.
    
    \textit{For a weak turbulence level:}
    At values $l_\mathrm{min}/(2\pi\,l_\mathrm{c})\lesssim \rho \lesssim l_\mathrm{min}/(\pi\,l_\mathrm{c}\,(b/B))$, particles scatter resonantly given appropriate pitch angles. As scattering is prohibited for parts of the pitch-angle spectrum, mirroring occurs instead around $\mu \approx 0$. Two effects oppositely affect the diffusion coefficient with reduced rigidity: the reduced range of allowed pitch angles enhances diffusion, while mirroring reduces parallel diffusion. Thus, the diffusion coefficient decreases somewhat above the boundary $\rho \sim r_\mathrm{g}/l_{\min}$ but then increases toward the upper end of the MR because the range of pitch angles that can scatter resonantly is widening until the resonant scattering regime is reached.
    
    \textit{For a strong turbulence level:} 
 	In the case of a weak or even absent background field, the direction of the magnetic field automatically provides for a changing $\mu$ along the particle's path due to the magnetic field's chaotic nature. Thus, the effect of missing resonant interactions towards low reduced rigidities is significantly attenuated and is only pronounced at gyroradii smaller than $l_\mathrm{min}/(2\pi)$ for $B = 0$. In addition, magnetic mirroring is not a dominant effect, since the magnetic moment is not conserved sufficiently long. Even though the $\mu \approx 0$ problem is absent for $b\gg B$, the frequency of resonant interactions decreases towards low reduced rigidities, and particles follow field lines as in the NRSR. 
    This effect is demonstrated in Fig.~\ref{fig:b_l_min}, where the diffusion coefficients are presented as functions of the reduced rigidity $\rho$ for different fluctuation ranges.

    \textit{Interpolation introduces a guide field even for $b\gg B$ on scales on the order of the grid spacing:}
    The linear interpolation algorithm of the magnetic field between grid points locally removes the turbulent character and therefore effectively introduces a guide field at scales on the order of the spacing of the grid points. If fluctuations extend towards these scales such that low-rigidity particles still scatter resonantly, the transport behaviour is similar to that for weak turbulence levels. This is demonstrated by the green triangles in Fig.~\ref{fig:b_l_min}.

    \item \textbf{Resonant Scattering Regime (RSR)}: Particles within the range $l_\mathrm{min}/(l_\mathrm{c}\pi(b/B)) \lesssim \rho \lesssim 1/(2\pi)$ can scatter resonantly over the complete range of the pitch angle as derived in the following. Individual particles scatter on average by $\delta \mu/\mu \approx b/B$ \citep{Kulsrud1969}. This effect is depicted in the two lower panels of Fig.~\ref{fig:influence_scatter} in the light blue area with white stripes, which represents the area into which particles are able to scatter on average. As soon as particles can, statistically, scatter across the gap around $\mu \approx 0$, they are not trapped anymore and mainly interact as described within QLT for $b \ll B$. The condition for particles to jump over the gap around $\mu\approx 0$ reads $\delta \mu \approx b/B \geq 2 \mu_\mathrm{min} =  l_\mathrm{min}/(\pi r_\mathrm{g})$. Consequently, the lower boundary of the RSR is determined by the minimal gyroradius
    \begin{equation}\label{eq:rho_min}
        \rho_\mathrm{min} = r_\mathrm{g,min}/l_\mathrm{c} = \frac{l_\mathrm{min}}{l_\mathrm{c}\pi(b/B)},
    \end{equation}
    for which the above condition still holds.
    Only within the RSR are the parallel-diffusion-coefficient dependencies expected to follow QLT, because the effective coverage of $\mu$ coincides with the total possible range $\Delta \mu/\mu_\mathrm{tot} = 1$. In addition, in order for particles to be influenced by fluctuations at certain mode numbers, the mode density must be sufficiently high \citep{Mace2012,Snodin2015}.
    \item \textbf{Transition Regime (TR)}: As soon as some particles cannot interact resonantly with fluctuations due to their value of $\mu$, the transition towards the quasi-ballistic regime begins. The lower boundary of this transition regime follows from the gyroresonance condition and yields $\rho = l_\mathrm{max}/(2\pi l_\mathrm{c})\approx5/(2\pi)$, independent of $l_\mathrm{min}$. With growing $\rho$, the percentage of particles that can still resonantly scatter decreases. \citet{Globus2007} estimates the range of this regime to be approximately one order of magnitude, $5/(2\pi) \lesssim \rho \lesssim 25/\pi$.
    \item \textbf{Quasi-Ballistic Regime (QBR)}: For particles with gyroradii $ 25/\pi \lesssim \rho$ that substantially exceed the correlation length of the turbulence, the transport behaviour converges toward ballistic propagation. This regime is called quasi-ballistic regime, because interactions of the particles around $\mu \approx 0$ are still possible according to the gyroresonance condition as illustrated in the two lower panels of Fig.~\ref{fig:influence_scatter}. The parallel-diffusion-coefficient dependency yields $\kappa_\parallel \propto (B/b)^2\rho^2cl_\mathrm{c}$, as derived in Sec.~\ref{Large_Rho}.
    In numerical simulations, it is important to consider the following:
    In the limit $\rho \gg 1$, the step size $s$ has to be chosen such that the magnetic field is still correlated at two subsequent particle positions: $s\ll l_\mathrm{c}$. Otherwise, the reduced-rigidity dependency of the diffusion coefficient $\kappa_\parallel \propto (B/b)^2 r_{\mathrm{g}}c$ is polluted as derived in Appendix~\ref{app:A}.
\end{enumerate}

The above five regimes with different diffusion coefficient dependencies are summarised in Tab.~\ref{tab:theoretical predictions} and illustrated in Fig.~\ref{fig:influence_scatter}. The upper panel of the figure schematically presents the expected dependencies of the parallel diffusion coefficient on reduced rigidity for a fixed range of fluctuations and turbulence levels $b/B$.

A key result to emerge from these considerations can be phrased as follows: The parallel diffusion coefficient greatly depends on the lower boundary $l_\mathrm{min}$ of the fluctuations, because this quantity determines the classification at a given $\rho$ for an otherwise fixed set of parameters. As soon as the diffusion coefficient is governed by the MR or NRSR instead of the RSR, its value increases. 

Fig.~\ref{fig:l_min} illustrates the latter argument by presenting parallel diffusion coefficients as functions of the reduced rigidity for different values of $l_\mathrm{min}$. The upper fluctuation boundaries $l_\mathrm{max}$ are adjusted as $l_\mathrm{min}$ is changed in such a way that the correlation length always has the same value $l_\mathrm{c}\approx 20$~pc, so that all curves coincide based on the theoretical considerations presented in Sec.~\ref{sec:2}. The diffusion coefficient for low reduced rigidities and for a given set of parameters converges toward its final, unpolluted value for decreasing $l_\mathrm{min}$. Consequently, an improperly high choice of $l_\mathrm{min}$ may result in artificially too high diffusion coefficients and subsequently in an artificially weak reduced-rigidity dependency. Thus, the diffusion coefficient may only converge to the value predicted within QLT\footnote{\citet{Schlickeiser1989} showed that the singularity in the quasi-linear diffusion coefficient can be removed if the finite-frequency effect is retained. That is, the resonance condition is $\omega - k \mu=\pm \Omega$; in the present paper $\omega$ is neglected implicitly, which may only be applicable for $v_A/c \lesssim \mu$, with $v_A$ being the Alfv{\'e}n speed \citep{Kulsrud1969}.} for small ratios $b/B$ and values of $l_\mathrm{min} \lesssim \pi \rho\,l_\mathrm{c} b/B$.

Although the short-wavelength cutoff is dictated by numerical considerations for the present case, in many astrophysical plasmas a similar threshold may exist due to strong damping processes at short wavelengths.

\begin{table*}
\caption{Definitions and ranges of the different scattering regimes and predicted parallel-diffusion-coefficient dependencies, as illustrated in Fig.~\ref{fig:influence_scatter}. }
\label{tab:theoretical predictions}
\begin{tabular}{cccccc}
	\hline 
	Regime & $l_\mathrm{min}/(2\pi r_\mathrm{g})$ &$l_\mathrm{max}/(2\pi r_\mathrm{g})$ &  $\rho$&  $b \ll B$ & $b \gg B$\tabularnewline
	\hline 
	NRSR & $ > 1$&$ \geq 1$ &$0-l_\mathrm{min}/(2\pi l_\mathrm{c})$ &$-$ & $-$ \tabularnewline
	MR & $ \napprox 0~ \&~ \leq 1$&$ \geq 1$ & $l_\mathrm{min}/(2\pi l_\mathrm{c})-l_\mathrm{min}/(\pi l_\mathrm{c} (b/B))$& $-$  & $-$\tabularnewline
	RSR & $ \approx 0$& $\geq 1$ &$l_\mathrm{min}/(\pi l_\mathrm{c}(b/B))-5/(2\pi)$  &$\kappa_\parallel \propto  \rho^{2-\alpha} c \,l_{\mathrm{c}} B^2/b^2$ &$\kappa_\parallel \propto cl_\mathrm{c}\rho B/(6b)$\tabularnewline
	TR & $\approx 0$&$<1$ &$5/(2\pi)-25/\pi$ & $-$ & $-$\tabularnewline 
	QBR & $\approx 0$&$\ll 1$ & $25/\pi-\infty$&$ \kappa_\parallel \propto (B/b)^2  \rho^2 c \,l_{\mathrm{c}} $ &$ \kappa_\parallel \propto (B/b)  \rho^2 c \,l_{\mathrm{c}} $ \tabularnewline
	\hline 
\end{tabular}
\end{table*}

\subsubsection{Consequences for Numerical Settings}\label{Numerical_Influence}
As demonstrated before, the diffusion coefficient converges toward its predicted value within QLT only if $l_\mathrm{min}$ is chosen sufficiently small. The required $l_\mathrm{min}$ depends on the turbulence level according to Eq.~(\ref{eq:rho_min}). However, to test QLT with numerical simulations not only requires one to resolve resonant scattering over the complete range of $\mu$, but also to fulfill additional conditions\footnote{These conditions are valid for regular grids. Whether they also have to be considered when using nested grids for the turbulence generation cannot be determined here. The box size and interpolation conditions are not required for the grid-free method that uses the superposition of plane waves for generating the fluctuations.}:
\begin{enumerate}
    \item \textbf{Box size:} Figure~\ref{fig:BoxSize} presents the final diffusion coefficient as a function of the box size. The colour indicates the CPU simulation time in arbitrary units. It shows that the grid must exceed a certain size before $\kappa_\parallel$ converges. Considering both requirements, small $l_\mathrm{min}$ and a large grid volume, a small lower limit of the fluctuation $l_\mathrm{min}$ is accompanied by a small spacing $s_\mathrm{spacing}$, in order to resolve of all fluctuations. For a large grid volume, the number of grid points has to be chosen correspondingly large, governed by $N_\mathrm{grid}\geq 2l_\mathrm{max}/s_\mathrm{spacing}$.
\begin{figure}
	\includegraphics[width=\columnwidth]{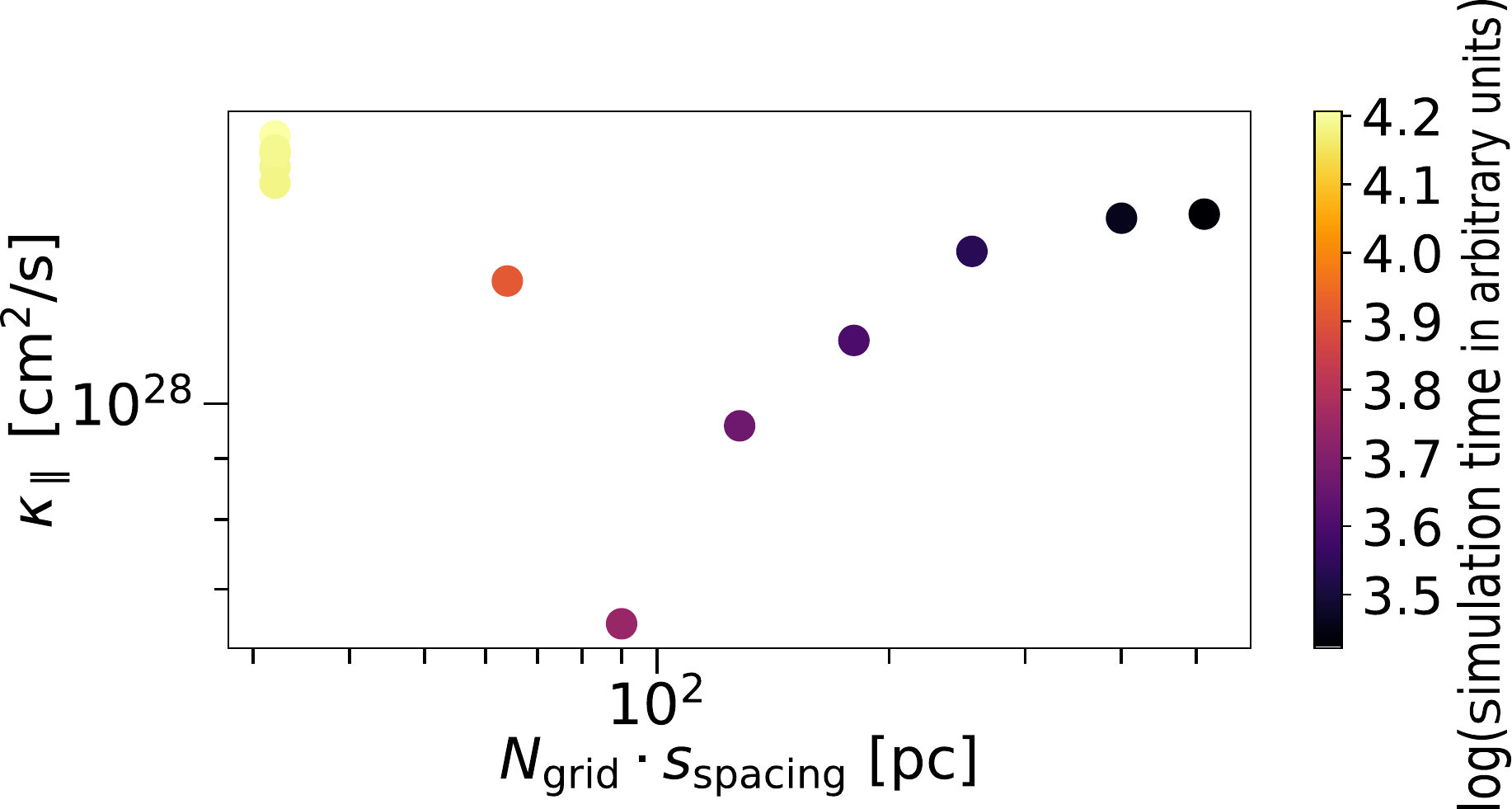}\caption{The parallel diffusion coefficient for different values of the product of the number of grid points with the spacing, which represents a measure of the grid size. Simulated with $l_\mathrm{min} = 1.7~\mathrm{pc}$,
		$l_\mathrm{max} = 82.5~\mathrm{pc}$, $b = 0.1~\mu\mathrm{G}$, and $B = 1~ \mu\mathrm{G}$. With more grid points, the generation of the turbulent field on the grid points takes longer. However, the CPU time of the magnetic field generation is only a small amount of the complete CPU simulation time in relation to the remaining CPU propagation time. Consequently, the number of grid points has no significant influence on the CPU simulation time.  }\label{fig:BoxSize}%
\end{figure}
\item \textbf{Step size:} Figure \ref{bp-ck-diff} presents the parallel diffusion coefficient as a function of the step size divided by the gyroradius. The final diffusion coefficient converges once the gyration motion is resolved sufficiently well. Since the Boris push is almost one order of magnitude faster than the Cash-Karp algorithm with the same precision, the Boris push is used for all subsequent simulations in this paper. A step size $s=r_\mathrm{g}/10$ is applied in all further simulations to guarantee high accuracy.
\item \textbf{Interpolation:} In trying to reproduce the QLT prediction in numerical simulation, it is vital to obey, for a sufficiently large RSR, $l_\mathrm{min} \leq r_\mathrm{g}\,\pi\,(b/B)$. In addition, it is helpful to increase $s_\mathrm{spacing}$ to increase the box size.
While it seems reasonable to decrease $l_\mathrm{min}/s_\mathrm{spacing}$ as much as possible ($l_\mathrm{min}/s_\mathrm{spacing} = 2$) to increase the energy range of the RSR, in \citet{Schlegel2019} it is demonstrated that this also increases the magnetic field interpolation error. The turbulence spectrum in the inertial range is artificially steepened because of the interpolation of the magnetic field between grid points. A ratio of $l_\mathrm{min}/s_\mathrm{spacing}=10$ comprises a good compromise of minimising the interpolation error and still allowing for a sufficient extension of the RSR.
\begin{figure}
	\includegraphics[width=\columnwidth]{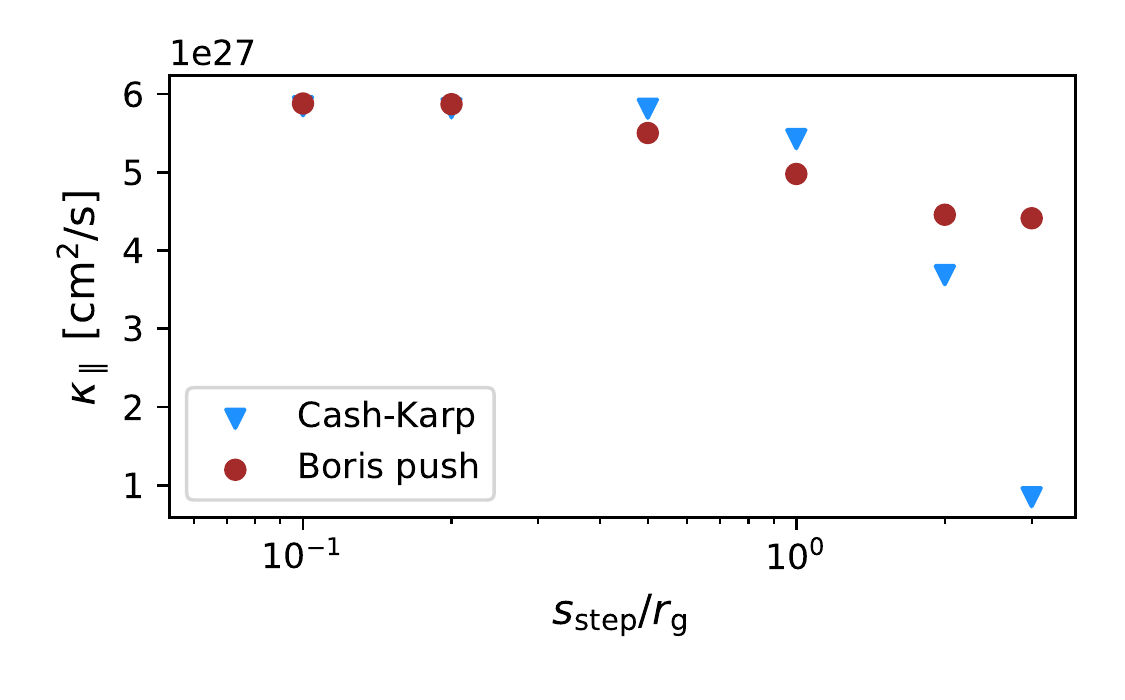}\caption{Comparison of both propagation methods -- the Cash-Karp algorithm and the Boris push -- with respect to the value of the diffusion coefficient as a function of the step length. 
	Too large a step size cannot resolve the particle motion with sufficient accuracy and will therefore pollute the diffusion coefficient. The numerically converged diffusion coefficient is only obtained when the step size is chosen small enough so that the gyration motion can be resolved.
	Simulated with $l_\mathrm{min} = 1.7~\mathrm{pc}$,
		$l_\mathrm{max} = 82.5~\mathrm{pc}$,
		$E = 8900$~TeV, $N_{\mathrm{grid}} = 1024$, $b = 0.1~\mu\mathrm{G}$, and $B = 1~\mu\mathrm{G}$.}\label{bp-ck-diff}%
\end{figure}
\item Magnetic-field realisations: The numerical calculation of turbulence using the grid method inevitably generates anisotropies due to the limited grid resolution. Instead of relying on only one of these randomly generated magnetic fields, simulations should be repeated for other field realisations. Therefore, to obtain isotropic turbulence that can serve as a realistic numerical set-up for comparison with theoretical predictions, averaging over many simulations with different random-phase realisations of the Kolmogorov turbulence using the same parameters is necessary \citep{Giacalone1999, Snodin2015}.
\end{enumerate}
Table \ref{paper} summarises key parameters used in previous studies, focussing on the ranges of different ratios $b/B$ and $r_\mathrm{g}/l_{\mathrm{c}}$. Listed are the range of wavenumbers along with the magnetic mode density.
The mode density per decade for simulations based on the wave model is defined as $N_\mathrm{m}/\log10(k_\mathrm{max}/k_\mathrm{min})$ due to the logarithmically spaced wavevectors, while $10\,(N_\mathrm{grid}/2)^3/(k_\mathrm{max}/k_\mathrm{min})$ is the definition of the mode density per decade for the grid-based turbulence method with linearly spaced wavevectors. $B_\mathrm{tot} = \sqrt{B^2+b^2}$ takes both magnetic field components into account. The gyroradii are calculated with respect to the background field unless stated otherwise. The ranges of reduced rigidities of the simulations are listed. The energy ranges can be rescaled as demonstrated in Appendix~\ref{app:B}. The upper boundary of the RSR is $\rho_\mathrm{max}=5/(2\pi)$. As discussed in Sec.~\ref{Numerical_Influence}, the numerical influence of the interpolation of the magnetic field on the diffusion coefficients depends on the ratio $l_\mathrm{min}/s_\mathrm{spacing}$. Instead of this ratio, the quantity $l_\mathrm{min}N_\mathrm{grid}/(4l_\mathrm{max})$ is presented, since the spacing between the grid points is not mentioned in most publications. For the turbulence generated on a grid, this value represents how well the generated waves fit into the grid, {{while averaging over many simulations with different realisations improves the effective isotropy of the field}}. A large ratio reduces the numerical effect introduced by interpolation \citet{Schlegel2019}.
The different power-law indices $\gamma$ of the energy dependency of the diffusion coefficient $\kappa_{\parallel}\propto E^{\gamma}$ are provided, as well. 

The simulation data between previous papers and this study differ only slightly and are consistent with each other. The difference in the resulting energy dependency is largely due to our restriction of the RSR according to the formalism established in this paper and the fact that we fit our simulation data, whereas in most previous papers only a match with QLT is indicated using a drawn line with slope $1/3$ at energies below the RSR. We expect that interpreting previous studies in light of the findings of this paper will result in similar values for $\gamma$ and consequently a turbulence-level-dependent energy scaling of $\kappa$.

\begin{table*}
    \centering
	\caption{
		Review of the physical and numerical input parameters used here and in previous studies. The different power-law indices $\gamma$ from the energy dependency of the diffusion coefficient $\kappa_{\parallel}\propto E^{\gamma}$ are listed for different ratios of $b/B$ as stated in each paper for Kolmogorov fluctuations. If agreement of the simulation results with QLT is found in the paper (even if no explicit fit is shown), a value of $1/3$ is listed in the last column ($2/3$ for non-relativistic particles \citep{Giacalone1999}).
		Publications use either a superposition of plane waves for generating the fluctuations or a discrete cubic grid. All models are based on isotropic turbulence. Key parameters of the simulations are quoted, such as the maximum fluctuation, the extent of the fluctuations $k_\mathrm{max}/k_\mathrm{min}$, and the number of modes $N_\mathrm{m}$ (wave model) or grid points $N_\mathrm{grid}$ along one direction (grid model). The maximum fluctuations $l_\mathrm{max}$ are quoted from the individual papers while, for reasons of comparability, the correlation lengths are uniformly computed according to the formula $l_\mathrm{c} = l_\mathrm{max}/5$ (see Eq.~(\ref{corr_l_max})). It is important to note, however, that $l_\mathrm{c}$ in some papers, marked with *, are calculated differently in the respective publications.}\label{paper}

	\begin{threeparttable}
	\scalebox{0.75}{
		\begin{tabular}{c|cccccccccccc}
			\hline 
			&$\frac{b}{B}$ &	$\array{c}\sqrt{B^2+b^2}\\\mathrm{[\mu G]}\endarray$	&$\array{c}l_\mathrm{max}\\\mathrm{[pc]}\endarray$	 &$\array{c}l_\mathrm{c}\\\mathrm{[pc]}\endarray$&	$\frac{k_{\mathrm{max}}}{k_{\mathrm{min}}}$&$\frac{l_\mathrm{min}}{l_\mathrm{max}}\frac{N_\mathrm{m/grid}}{4}$&$\array{c}\mathrm{mode}\\\mathrm{density}\endarray$&	$N_\mathrm{m/grid}$&	$\rho$&	$\array{c}\mathrm{turbulence}\\\mathrm{model}\endarray$ &		$\gamma$   ~\tabularnewline
			\hline 
			\hline

			present paper& $1.49$ &1.79&	82.5& {{$17$}}&	48.5 &5.3&	$2.8\cdot 10^7$&	1024&	$0.02-7.87$&grid&$0.987\pm0.009$  \tabularnewline
			
			present paper& $1.17$ &1.54&	82.5&{{$17$}}&	48.5 &5.3&	$2.8\cdot 10^7$&	1024&	$0.02-7.87$&grid&$1.002\pm0.005$  \tabularnewline
			
			present paper& $0.92$ &1.36&	82.5&{{$17$}}&	48.5 &5.3&	$2.8\cdot 10^7$&	1024&	$0.02-7.87$&grid&$0.990\pm0.010$  \tabularnewline
			
			present paper& $0.73$ &1.24&	82.5&{{$17$}}&	48.5 &5.3&	$2.8\cdot 10^7$&	1024&	$0.02-7.87$&grid&$0.966\pm0.012$  \tabularnewline
			
			present paper& $0.57$ &1.15&	82.5&{{$17$}}&	48.5 &5.3&	$2.8\cdot 10^7$&	1024&	$0.02-7.87$&grid&$0.934\pm0.011$  \tabularnewline
			
			present paper& $0.45$ &1.10&	82.5&{{$17$}}&	48.5 &5.3&	$2.8\cdot 10^7$&	1024&	$0.02-7.87$&grid&$0.882\pm0.011$  \tabularnewline
			
			present paper& $0.36$ &1.06&	82.5&{{$17$}}&	48.5 &5.3&	$2.8\cdot 10^7$&	1024&	$0.02-7.87$&grid&$0.848\pm0.014$  \tabularnewline
			
			present paper& $0.28$ &1.04&	82.5&{{$17$}}&	48.5 &5.3&	$2.8\cdot 10^7$&	1024&	$0.02-7.87$&grid&$0.815\pm0.015$  \tabularnewline
			
			present paper& $0.22$ &1.02&	82.5&{{$17$}}&	48.5 &5.3&	$2.8\cdot 10^7$&	1024&	$0.02-7.87$& grid &$0.770\pm0.012$  \tabularnewline
			
			present paper& $0.17$ &1.02&	82.5&{{$17$}}&	48.5 &5.3&	$2.8\cdot 10^7$&	1024&	$0.02-7.87$&grid&$0.739\pm0.014$  \tabularnewline
			
			present paper& $0.14$ &1.01&	82.5&{{$17$}}&	48.5 &5.3&	$2.8\cdot 10^7$&	1024&	$0.02-7.87$&grid&$0.712\pm0.011$  \tabularnewline
			
			present paper& $0.11$ &1.01&	82.5&{{$17$}}&	48.5 &5.3&	$2.8\cdot 10^7$&	1024&	$0.02-7.87$&grid&$0.671\pm0.013$  \tabularnewline
			
			present paper& $0.09$ &1.00&	82.5&{{$17$}}&	48.5 &5.3&	$2.8\cdot 10^7$&	1024&	$0.02-7.87$&grid&$0.645\pm0.011$  \tabularnewline
			
			present paper& $0.07$ &1.00&	82.5&{{$17$}}&	48.5 &5.3&	$2.8\cdot 10^7$&	1024&	$0.02-7.87$&grid&$0.616\pm0.016$  \tabularnewline

			\hline
			
			\citet{Giacinti2017}& $\infty$ &1&	100&{{$20$}}&	64\tnote{$+$} &1&	$3.3\cdot 10^5$&	256&	$0.054-	5.405$\tnote{$\dagger$}&grid&$1/3$\tabularnewline
			
			\citet{Giacinti2017}& $4$ &1&	100&{{$20$}}&	64\tnote{$+$} &1&	$3.3\cdot 10^5$&	256&	$0.054-	5.405$\tnote{$\dagger$}&grid&$1/3$ \tabularnewline
			
			\citet{Giacinti2017}& $2$ &1&	100&{{$20$}}&	64\tnote{$+$} &1&	$3.3\cdot 10^5$&	256&	$0.054-	5.405$\tnote{$\dagger$}&grid&$1/3$ \tabularnewline
			
			\citet{Giacinti2017}& $1$ &1&	100&{{$20$}}&	64\tnote{$+$} &1&	$3.3\cdot 10^5$&	256&	$0.054-	5.405$\tnote{$\dagger$}&grid&$1/3$  \tabularnewline
			
			\citet{Giacinti2017}& $0.5$ &1&	100&{{$20$}}&	64\tnote{$+$} &1&	$3.3\cdot 10^5$&	256&	$0.054-	5.405$\tnote{$\dagger$}&grid&$1/3$  \tabularnewline
			
			\citet{Giacinti2017}& $0.1$ &1&	100&{{$20$}}&	64\tnote{$+$} &1&	$3.3\cdot 10^5$&	256&	$0.054-	5.405$\tnote{$\dagger$}&grid&$1/3$  \tabularnewline
		
		    \hline
		    
			\citet{Subedi2017}& $\infty$& $1$ &  $-$ &{{$-$\tnote{$*$}}} & $-$ &{{$-$}} & $-$&1024&$0.001-  20$\tnote{$\dagger$} & grid &1/3 \tabularnewline
			
			\hline 
			
			\citet{Snodin2015} &	$\infty$		&		$-$&$-$\tnote{$\ddagger\ddagger$}	&{{$-$\tnote{$*$}}}&384~/~256&1~/~0.5&				$1.2 \cdot 10^7~$/$~213$&	$1536~$/$~512$&	$0.01-	2.5	$\tnote{$\dagger$}&grid $/$ waves&				1 
			\tabularnewline
			
			\citet{Snodin2015} &	9.95		&		$-$ &	$-$\tnote{$\ddagger\ddagger$}&{{$-$\tnote{$*$}}}&200 &	1.3&			445&	1024&	$0.01-	2.5	$\tnote{$\dagger$}&waves&		1/3	
			\tabularnewline
			
			\citet{Snodin2015} & 3		&		$-$ &$-$\tnote{$\ddagger\ddagger$}&{{$-$\tnote{$*$}}}&200 &1.3&				445&	1024&	$0.01-	2.5	$\tnote{$\dagger$}&waves&		1/3		
			\tabularnewline
			
			\citet{Snodin2015} &	1		&		$-$&	$-$\tnote{$\ddagger\ddagger$}&{{$-$\tnote{$*$}}}&200 &1.3&				445&	1024&	$0.01-	2.5	$\tnote{$\dagger$}&waves&		1/3		
			\tabularnewline
			
			\citet{Snodin2015} &	0.33		&		$-$ &	$-$\tnote{$\ddagger\ddagger$}&{{$-$\tnote{$*$}}}&200 &1.3&				445&	1024&	$0.01-	2.5$\tnote{$\dagger$}&waves&	1/3				
			\tabularnewline
			
			\hline 			
			
			\citet{Harari2013}& $\infty$& 0.01 &  $10^6$  &{{$2\cdot 10^5$}}& $\geq 50$& $-$& $-$&$-$&$0.0054-  54$\tnote{$\dagger$}&waves& 1/3 \tabularnewline
			
			\hline

			\citet{Fatuzzo2010}& {{$\infty$}}& {{$10$}} & {{$1$}}&{{$0.2$\tnote{$*$}}}& {{$10^4~/~10^5$ }}& {{$3\cdot10^{-4}-2.5\cdot 10^{-3}$}} &{{$25$}}&{{$100 ~/~125$}}&{{$0.0005-0.5$}} & {{waves}}&{{$1/3$}}\tabularnewline
			
			\citet{Fatuzzo2010}& {{$0.92$}}& {{$14.14$}} & {{$1$}}&{{$0.2$\tnote{$*$}}}& {{$10^4~/~10^5$ }}& {{$3\cdot10^{-4}-2.5\cdot 10^{-3}$}} &{{$25$}}&{{$100 ~/~125$}}&{{$0.0005-0.5$}}  & {{waves}}&{{$1/3$}}\tabularnewline

			\hline
			
			\citet{Globus2007}& $\infty$& 0.01 &  $10^6$ &{{$2\cdot 10^5$}}& $-$ &$-$& $-$&$-$&$0.0054-  54$\tnote{$\dagger$} & waves&1/3\tabularnewline

			\hline 
			\citet{DeMarco2007}& $\infty$	&100	&			100&{{$20$}}&	64\tnote{$+$} 		&1&	$3.3\cdot 10^5$	&256	&$0.054-5.405	$		&grid	&$-$ 	 \tabularnewline
			
			\citet{DeMarco2007}&2	&2.236	&			100&{{$20$}}&	64\tnote{$+$} 	& 1 &	$3.3\cdot 10^5$	&256	&$0.054-5.405$			&grid	&1/3 \tabularnewline
			
			\citet{DeMarco2007}&1	&1.414	&			100&{{$20$}}&	64\tnote{$+$} 	& 1 & $3.3\cdot 10^5$	&256	&$0.054-5.405	$			&grid	&1/3\tabularnewline
			
			\citet{DeMarco2007} &0.5	&1.118	&			100&{{$20$}}&	64\tnote{$+$}			&1&$3.3\cdot 10^5$	&256	&$0.054-5.405$	&grid	&1/3 \tabularnewline
			
			\hline

			\citet{Candia2004}&	10 &	10.05	&			100&{{$20$}}&	$10-10^3$	&$0.025-2.5$&			$33-100$&100	&$0.05-	5$	&waves&	1/3
			\tabularnewline
			
			\citet{Candia2004} &	 7.07&	7.14	&			100&{{$20$}}&	$10-10^3$	&$0.025-2.5$&			$33-100$	&100	&$0.05-	5$	&waves&	1/3
			\tabularnewline
			
			\citet{Candia2004} &	 5.48&	5.57	&			100&{{$20$}}&	$10-10^3$	&$0.025-2.5$&			$33-100$	&100	&$0.05-	5$	&waves&	1/3
			\tabularnewline
			
			\citet{Candia2004} &	 3.16&	3.31	&			100&{{$20$}}& $10-10^3$	&$0.025-2.5$&			$33-100$	&100	&$0.05-	5$	&waves&	1/3			\tabularnewline
			
			\citet{Candia2004} &	 2.24&	2.45	&			100&{{$20$}}&	$10-10^3$	&$0.025-2.5$&			$33-100$	&100	&$0.05-	5$	&waves&	1/3
			\tabularnewline
			
			\citet{Candia2004} &	 1.73&	2.00	&			100&{{$20$}}&	$10-10^3$	&$0.025-2.5$&			$33-100$	&100	&$0.05-	5$	&waves&	1/3
			\tabularnewline
			
			\citet{Candia2004} &	 1&	1.41	&			100&{{$20$}}&	$10-10^3$	&$0.025-2.5$&			$33-100$	&100	&$0.05-	5$	&waves&	1/3
			\tabularnewline
			
			\citet{Candia2004} &	 0.71&	1.23	&			100&{{$20$}}&	$10-10^3$	&$0.025-2.5$&			$33-100$	&100	&$0.05-	5$	&waves&	1/3
			\tabularnewline
			
			\citet{Candia2004} &	 0.55&	1.14	&			100&{{$20$}}&	$10-10^3$	&$0.025-2.5$&			$33-100$	&100	&$0.05-	5$	&waves&	1/3
			\tabularnewline
			
			\citet{Candia2004} &	 0.32&	1.05	&			100&{{$20$}}&	$10-10^3$	&$0.025-2.5$&			$33-100$	&100	&$0.05-	5$	&waves&	1/3
			\tabularnewline
			
			\hline
			\citet{Parizot2004} & $\infty$& 0.01 &  $10^6$&{{$2\cdot 10^5$}} & $10^2$ / $10^4$ &$ 2.5$ &$100$&$200$ / $400$&$0.0054$ / $54$ \tnote{$\dagger$}&waves& 1/3 \tabularnewline
			
			\hline
			
			\citet{Casse2001}&$\infty$ &$-$ & $-$\tnote{$\ddagger\ddagger$}&{{$-$\tnote{$*$}}}& 64\tnote{$+$} 	&1	&		$3.3\cdot10^5$&	256 &$	8\cdot10^{-5}-	5$\tnote{$\dagger$}		&grid&			1  \tabularnewline
			
			\citet{Casse2001}&9.95&			$-$& $-$\tnote{$\ddagger\ddagger$}& {{$-$\tnote{$*$}}}&64\tnote{$+$} &1&			$3.3\cdot10^5$&	256&	$0.006-	5		$\tnote{$\dagger$}&grid&			1/3 \tabularnewline
			
			\citet{Casse2001}&0.92&			$-$	&$-$\tnote{$\ddagger\ddagger$} &{{$-$\tnote{$*$}}}& 64\tnote{$+$} 	&1&			$3.3\cdot10^5$&	256&	$0.006-	5		$\tnote{$\dagger$}&grid&			1/3 \tabularnewline
			
			\citet{Casse2001}&0.52&			$-$	&$-$ \tnote{$\ddagger\ddagger$}&{{$-$\tnote{$*$}}}& 64\tnote{$+$} &1	&			$3.3\cdot10^5$&	256&	$0.006-	5		$\tnote{$\dagger$}&grid&			1/3\tabularnewline
			
			\citet{Casse2001}&0.33&		$	-$	&$-$ \tnote{$\ddagger\ddagger$}& {{$-$\tnote{$*$}}}&64\tnote{$+$} 	&1&			$3.3\cdot10^5$&	256&	$0.006-	5		$\tnote{$\dagger$}&grid&			1/3 \tabularnewline
			
			\hline

			\citet{Giacalone1999}& 1 & 70.71 & $2.4\cdot 10^{-7}$ &{{$0.5\cdot 10^{-8}$\tnote{$*$}}}& $10^4$&$-$ &$-$&$-$ &{{$0.001 - 0.04$}}  &waves& 2/3 
			\tabularnewline
			
			\hline

		\end{tabular}
		}
		\begin{tablenotes}\footnotesize 
		    \item[$+$] These parameters are converted according to the definitions in the present paper (see for example Eqs.~(\ref{l_cond1})) and (\ref{eq:spectrum})).
		    \item[$*$] {{The correlation length is defined differently in the cited paper. Here, the correlation length is uniformly calculated according to $l_\mathrm{c} = l_\mathrm{max}/5$ \\(see Eq.~(\ref{correlation_Length})) to ensure comparability of the different simulations. The range of reduced rigidity in the table is determined based on this \\uniformly-defined correlation length and may therefore deviate from the values presented in the papers.}}   
			\item[$\dagger$] This reduced-rigidity range is based on the definition $r_\mathrm{g}\propto 1/\sqrt{B^2+b^2}$, which is used in the quoted paper.
			\item[$\ddagger\ddagger$] Even if the concrete values for these parameters are missing, the boundaries of the regimes can be calculated according to the\\ formulas mentioned in Sec.~\ref{sec:3.1.1} and especially in Fig.~\ref{fig:influence_scatter}, taking into account the relation $l_\mathrm{c} \approx  \,l_\mathrm{max}/ 5$ so that for example the upper NRSR\\ boundary reads $\rho \approx r_\mathrm{g}/ l_\mathrm{c} \approx 5\,r_\mathrm{g}/ l_\mathrm{max}$.
			The lack of this information, however, prevents the calculation of the minimum fluctuation wavelength and thus\\ the ratio of $l_\mathrm{min}/s_\mathrm{spacing}$, which is an indicator of the interpolation effect (see Sec.~\ref{Numerical_Influence}).
		\end{tablenotes}
    \end{threeparttable}
    
\end{table*}

\section{Comparison of Reduced-Rigidity Dependencies Between Simulations and QLT}\label{sec:4}
\begin{figure}
	\includegraphics[width=\columnwidth]{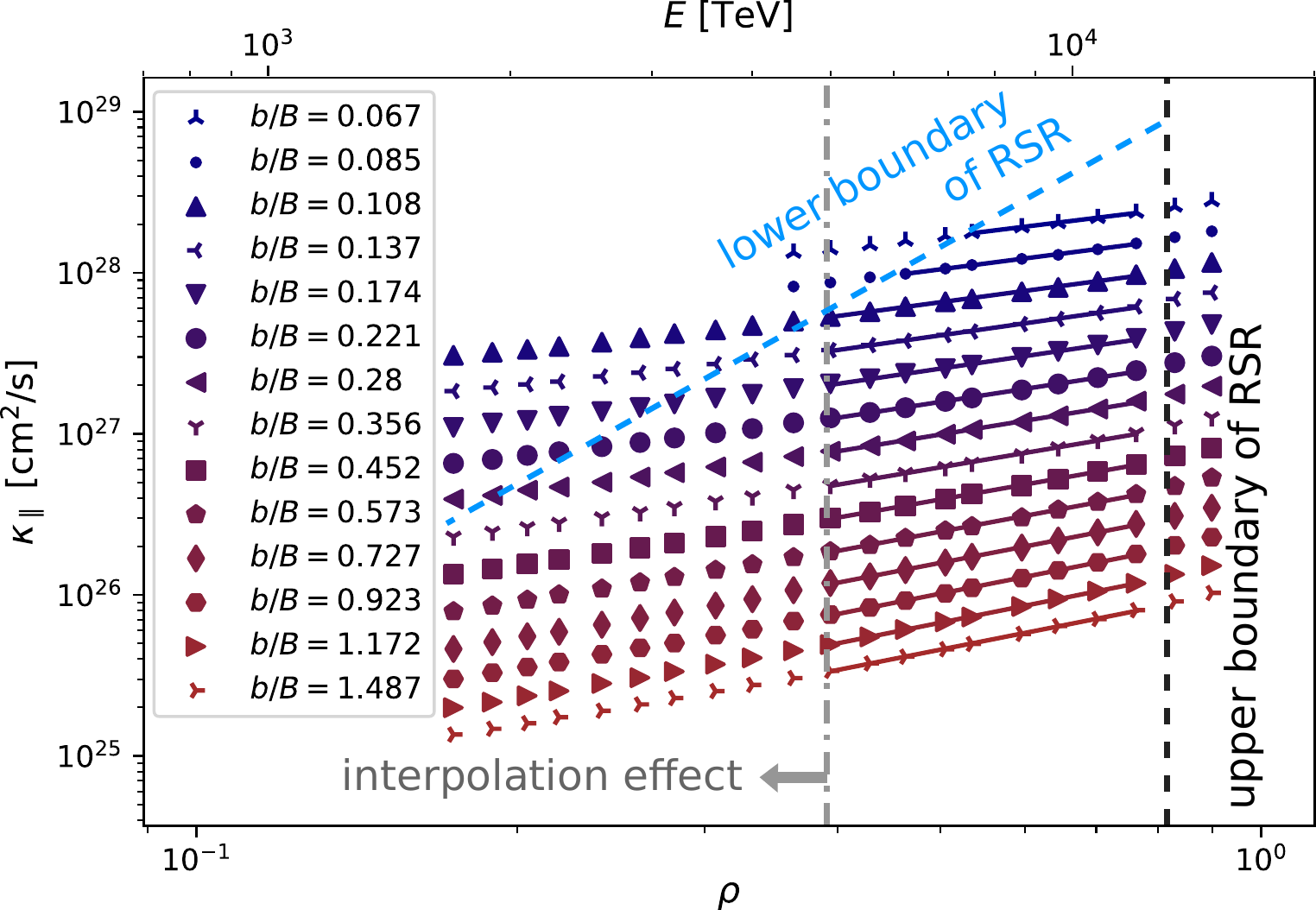}\caption{Parallel diffusion coefficients as functions of $\rho$ and $E$ for different turbulence levels. Only simulation results above the lower boundary of the RSR $\rho \gtrsim l_\mathrm{min}/(\pi(b/B)l_\mathrm{c})$ (light blue dashed line), above $\rho \gtrsim 0.3$ (grey dash-dotted line), where there is no noticeable effect of interpolation, and below the upper boundary of the RSR $\rho \lesssim 5/(2\pi)$ are considered for determining the energy scaling of $\kappa_\parallel$ within RSR. Fits of the equation from the QLT prediction $\kappa_\parallel = a\cdot \rho^{\gamma}$ are performed to these simulated diffusion coefficients, where $a$ is the proportional constant and $\gamma$ the power-law index.
     The parameters are $l_\mathrm{min} = 1.7$ pc, $l_\mathrm{max} = 82.5$ pc, $s = 0.17$ pc, $N_\mathrm{grid}$ = 1024. Each presented data point is the mean of 20 diffusion coefficients, each simulated with the same parameters but with a different turbulent field realisation. The decreasing range of the RSR for smaller $b/B$ leads to an increasing error in the slopes of the fits.}\label{fig:kappa_zz_diff_detailed}%
\end{figure}

This Section utilises the systematic developed in Sec.~\ref{sec:3} to evaluate the dependencies of the diffusion coefficients in the RSR numerically. We have applied our simulation results only to highly-relativistic protons, but the presented data can be re-scaled to other contexts. Figure~\ref{fig:kappa_zz_diff_detailed} presents diffusion coefficients calculated using 5000 particles in each simulation for 14 different ratios of $b/B$, where $B=1\,\mathrm{\mu G}$ was kept constant (the strength of $B$ is set for scaling purposes only and is not meant to correspond to a particular physical system). For each of these ratios, up to 21 different energies are simulated. Each data point is composed of 20 statistically independent simulations with the same parameters but different random-phase realisations of the Kolmogorov turbulence. The mean values are shown as functions of the reduced rigidity in Fig.~\ref{fig:kappa_zz_diff_detailed} together with their statistical uncertainties, which are, however, only a few percent and therefore too small to be visible.

The turbulence-level-dependent energy scaling of the diffusion coefficients is fitted to the data in the RSR. In addition to the physical boundaries of the RSR, the interpolation effect is considered for constraining the reduced-rigidity range of the fits: As pointed out in Sec.~\ref{Numerical_Influence}, the numerical error of the magnetic field interpolation increases toward low energies. Due to the high statistical accuracy of each individual data point, only a few points are necessary for each fit. A cut at $\rho \lesssim 0.3$ guarantees a sufficiently low influence of the interpolation routine, while accounting for a large enough range in energy to demonstrate linear behaviour in the log-log representation with low uncertainties. The deviation from a power-law energy scaling of $\kappa_\parallel$ toward low energies below the interpolation-effect cut-off is mainly caused by the magnetic field interpolation. As demonstrated in \citet{Schlegel2019}, the interpolated spectrum steepens (larger slope $\alpha$) toward small scales that are important for resonant scatterings with low-energy particles. This flattens the energy scaling of the diffusion coefficient according to $\kappa_\parallel \propto E^{2-\alpha}$, assuming an energy scaling consistent with QLT.

\begin{figure}
	\includegraphics[width=\columnwidth]{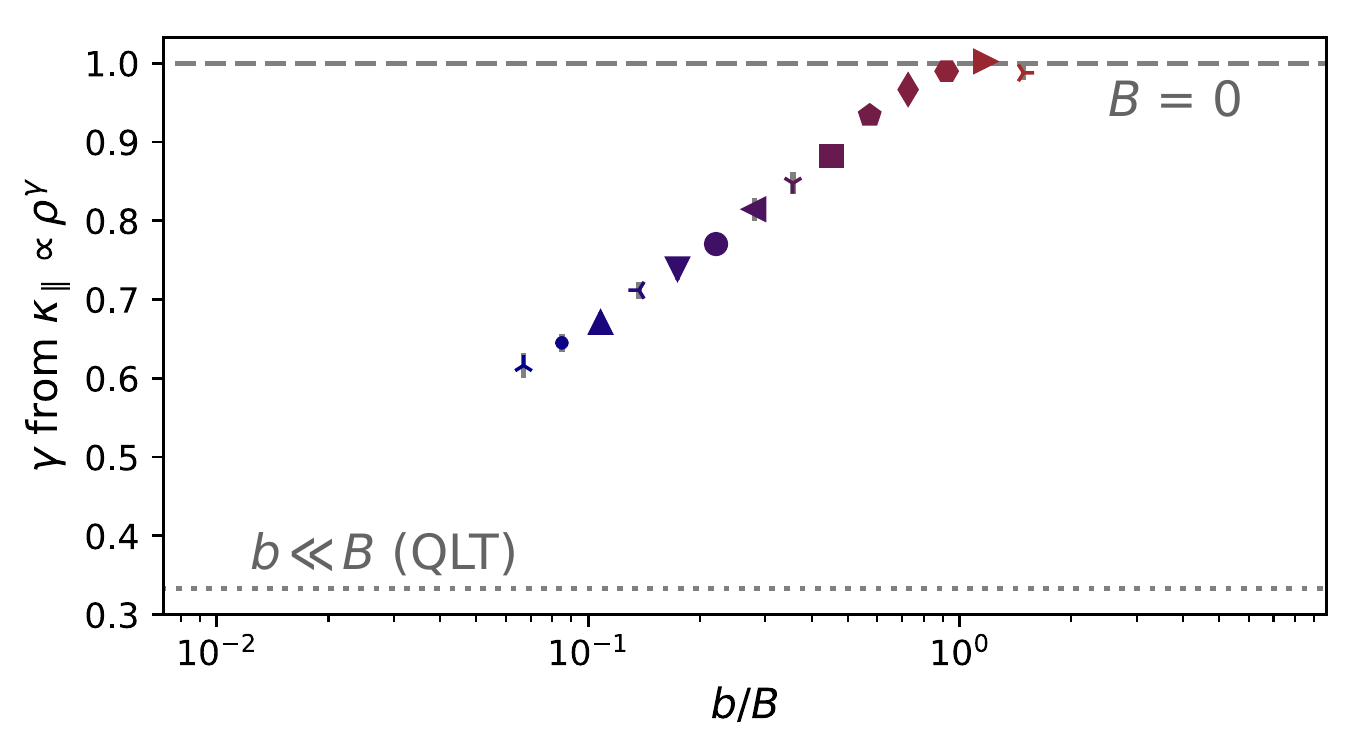} \caption{Turbulence-level-dependent spectral index of the diffusion coefficient in the RSR with (near-invisibly small) statistical errors. The simulated diffusion coefficients are fitted for each ratio of $b/B$ as shown in Fig.~\ref{fig:kappa_zz_diff_detailed}, with the slopes $\gamma$ shown here. The simulation parameters are $l_\mathrm{min} = 1.7\,$pc, $l_\mathrm{max} = 82.5\,$pc, $s = 0.17\,$pc, $N_\mathrm{grid}$ = 1024. {{The markers were chosen so that they correspond directly to those in Fig.~\ref{fig:kappa_zz_diff_detailed}}}.}\label{fig:summary_zz}
\end{figure}

Given the small error bars and the good quality of the fit, it can be inferred that at least locally in $\rho$, a power-law dependency clearly exists, and one may conjecture that under the right conditions, this dependency will extend over a much larger range. However, a physical situation where the RSR spans multiple orders of magnitude will require a presently unfeasibly costly numerical effort to resolve. Only once significantly more computing power is available will a direct test be possible whether this slope is representative of a greatly expanded RSR.
\\\\
One of the limitations of the theoretical predictions for the diffusion coefficient dependencies is that they were derived for the limit $b \ll B$ or $B = 0$. The expected values of $\gamma$ that are predicted from theoretical considerations are indicated with horizontal dashed lines in Fig.~\ref{fig:summary_zz}, where $\gamma$ is defined as the exponent of the power law
\begin{align}
    \kappa \propto \rho^{\gamma}.
\end{align}
Figure~\ref{fig:summary_zz} shows the $\gamma$ that results from the fits presented in Fig.~\ref{fig:kappa_zz_diff_detailed} as a function of the turbulence level.  
Even though the presented ratios of $b/B$ are still not small enough to agree with QLT predictions, a clear trend is visible: decreasing
$b/B$ decreases the slope, and the trend appears to be consistent with a value of $1/3$ for infinitesimal $b/B$, although the limit $b\ll B$ required for the QLT has not yet been reached in Fig.~\ref{fig:summary_zz}. In particular, our simulations reach down to turbulence levels $b/B\sim 0.05$, where the index is $\gamma\sim 0.6$ and thus still far from the value expected in QLT for highly relativistic particles. Thus, we can quantify three conclusions: (1) the limit of QLT is only valid for turbulence levels $b/B<0.05$; (2)  a turbulence-dependent diffusion coefficient is needed for the description of the parallel transport; (3) for the Bohm limit (dominating turbulence) the parallel diffusion coefficient converges toward a slope of one, as expected from Eq.~(\ref{large_b}).

\section{Discussion and outlook\label{discussion:sec}}
{{We have investigated, by means of direct numerical simulations, how cosmic rays of different energies diffuse in turbulent magnetic fields.}}
The two key findings of this work are
\begin{enumerate}
    \item The energy range for numerical simulations of diffusive propagation is highly constrained. 
    In a situation, where a simulation covers the entire wavevector spectrum with a physical $k_{\min}$ and $k_{\max}$, 
    the five regimes we present are physical and need to be considered in cosmic-ray propagation. It should be noted that our interpretation of the regions below and above the resonant scattering regime can change if we avoid sharp cutoffs in the wavevector spectrum [Eq.\ (\ref{eq:spectrum})]. In particular, in the mirroring regime, more waves for scattering will be available and the effect in the MR will be reduced and only become more prominent toward the boundary of the NRSR. Conclusions about the RSR, NRSR and QBR remain unchanged. In particular, our results pertaining to the diffusion coefficient are unaffected.
    \item By selecting an appropriate range for the fits to the energy dependence of the particles, we quantitatively show for the first time that QLT is \textit{not} valid at turbulence levels $b/B>0.05$ for Kolmogorov turbulence as can be seen in Fig.~\ref{fig:summary_zz}.
    Around $b/B\approx 1$, the Bohm diffusion limit $\kappa\propto \rho$ is reached. Qualitatively, the steeper energy dependence of the diffusion coefficient at larger $b/B$ occurs because higher energy particles ``see’’ the larger-amplitude turbulence first and start transitioning to the Bohm regime before lower-energy particles do. A more quantitative explanation of this effect is beyond the scope of this paper and will be addressed in future work. Although this work has focused on the energy range in this context, for other applications one may base analyses on the more fundamental reduced rigidity.
\end{enumerate}
These results can be put into an astrophysical context, specifically diffuse gamma-ray emission from the Milky Way. A radial gradient exists in the proton spectral index observed in the Galaxy \citep{fermi2016} --- the cosmic-ray spectrum in the central molecular zone, i.e., the inner 200 pc, is very flat, with $dN/dE\propto E^{-2.3}$. At a radius of $0.2-1.5$~kpc from the Galactic Center, the spectrum becomes extremely steep, $E^{-3.1}$, then reflattening to about $E^{-2.6}-E^{-2.7}$ up to $8$~kpc. In the outskirts of the Galaxy at $>8$~kpc, the spectrum becomes steeper again with $E^{-2.8}-E^{-2.9}$; compare \citet{yang2016}.

Cosmic-ray self-confinement via the streaming instability \citep{Kulsrud1969} has an influence on the spectrum. However, this requires cosmic-ray energies below which the cosmic-ray flux, which excites the instability, is large enough to overcome damping by the thermal background. It has long been recognised that above this critical energy, there must be a transition to confinement by turbulence from another source. Estimates for this critical energy are in the $100 - 300 \, \mathrm{GV}$ range, depending on the damping mechanism \citep{Cesarsky1973, Farmer2004,blasi2012}. This could produce a spectral break as observed in cosmic-ray data \citep{blasi2012,evoli2019}. It is unclear, however, if the instability can be maintained up to energies as high as 100 GV \citep{schlickeiser2016}. It is also beyond the scope of this work to tie this to a trend with galactocentric radius, and we simply point out that such an influence needs to be taken into account for a full simulation of Galactic propagation.

Galactocentric effects that could cause the steepening in the spectrum can be divided into data reduction problems and transport-related phenomena. We provide a list and argue that our present findings support argument number 5:

\begin{enumerate}
    \item Unresolved point sources could play a role. While \citet{pothast2018} argues that this contribution should be negligible, its role is not fully understood \citep{grenier_privcom}.
    \item A limited understanding of the gas distribution, and with it a possible systematic error in the data, cannot be excluded. This is particularly true for the central volume with $r<1$~kpc \citep{fermi2016}, which could have steeper cosmic-ray spectra. However, data at TeV energies exist indicating that the local spectrum is quite flat \citep{hess_gc_2016}.
    \item A Galactic wind keeps the spectral behaviour of observed cosmic rays constant at the level of injection. This would explain the observed flat component in the central molecular zone, assuming dominance of the wind in the Galactic Center region \citep{gaggero_prl2017,pothast2018}. 
    \item A geometric effect of different orientations of the total magnetic field along the galactocentric radius \citep{gaggero2015} could contribute to the gradient.
    \item Deviations from Kolmogorov-type diffusion in QLT have been discussed \citep{gaggero2015}. A radial dependence of the spectral index of the diffusion coefficient in the Galaxy has been proposed to explain the spectral softening toward the outer parts of the Galaxy \citep{gaggero_prl2017}, i.e., $\alpha=B+A r$. 
\end{enumerate}
This last effect, which has commonly been employed as a phenomenological explanation \citep{gaggero2015,gaggero_prl2017}, can now be supported by fundamental arguments: The turbulence level increases toward the outer parts of the Galaxy \citep{Jannson2012, kleimann2019, shukurov2019}. With the increase of the diffusion spectral index toward higher turbulence levels, we expect the spectrum toward large galactocentric radii to become steeper. Our results indicate that the scenario of a diffusion-driven change in the spectral index needs to be taken into account when trying to explain the cosmic-ray gradient problem in the Galaxy. Future work on detailed simulations of Galactic transport, including the $b/B$ dependence as derived here, in comparison with state-of-the-art observations will help to discriminate the different scenarios.


\section*{Data Availability}
Simulations were performed with the publicly available tool CRPropa \citep{AlvesBatista2016} (the specific version used for the simulations is CRPropa 3.1-f6f818d36a64), supported by various analysis tools \citep{Hunter_2007,pandas,van_der_Walt_2011,jupyter-notebook,Virtanen2019}. The data analysed in this article can be made available upon reasonable request to the corresponding author. 


\section*{Acknowledgements}

We would like to thank P. Desiati, A. Dundovic, H. Fichtner, R. Grauer, G. Giacinti, I. Grenier, R. Schlickeiser, A. Shalchi and A. Shukurov for highly valuable discussions. We acknowledge support from U.S.~DOE grant DE-FG02-04ER-54742 (MJP) and NSF grant AST 1616037 (EGZ). This work is supported by the ``ADI 2019’’ project funded by the IDEX Paris-Saclay, ANR-11-IDEX-0003-02 (PR).




\bibliographystyle{mnras}
\bibliography{source} 

\begin{thebibliography}{}
\makeatletter
\relax
\def\mn@urlcharsother{\let\do\@makeother \do\$\do\&\do\#\do\^\do\_\do\%\do\~}
\def\mn@doi{\begingroup\mn@urlcharsother \@ifnextchar [ {\mn@doi@}
  {\mn@doi@[]}}
\def\mn@doi@[#1]#2{\def\@tempa{#1}\ifx\@tempa\@empty \href
  {http://dx.doi.org/#2} {doi:#2}\else \href {http://dx.doi.org/#2} {#1}\fi
  \endgroup}
\def\mn@eprint#1#2{\mn@eprint@#1:#2::\@nil}
\def\mn@eprint@arXiv#1{\href {http://arxiv.org/abs/#1} {{\tt arXiv:#1}}}
\def\mn@eprint@dblp#1{\href {http://dblp.uni-trier.de/rec/bibtex/#1.xml}
  {dblp:#1}}
\def\mn@eprint@#1:#2:#3:#4\@nil{\def\@tempa {#1}\def\@tempb {#2}\def\@tempc
  {#3}\ifx \@tempc \@empty \let \@tempc \@tempb \let \@tempb \@tempa \fi \ifx
  \@tempb \@empty \def\@tempb {arXiv}\fi \@ifundefined
  {mn@eprint@\@tempb}{\@tempb:\@tempc}{\expandafter \expandafter \csname
  mn@eprint@\@tempb\endcsname \expandafter{\@tempc}}}

\bibitem[\protect\citeauthoryear{{Acero} et~al.,}{{Acero}
  et~al.}{2016}]{fermi2016}
{Acero} F.,  et~al., 2016, \mn@doi [\apj] {10.3847/0067-0049/223/2/26}, \href
  {https://ui.adsabs.harvard.edu/abs/2016ApJS..223...26A} {223, 26}

\bibitem[\protect\citeauthoryear{{Adhikari}, {Zank}, {Hunana}, {Shiota},
  {Bruno}, {Hu}  \& {Telloni}}{{Adhikari} et~al.}{2017}]{Adhikari2017}
{Adhikari} L.,  {Zank} G.~P.,  {Hunana} P.,  {Shiota} D.,  {Bruno} R.,  {Hu}
  Q.,   {Telloni} D.,  2017, \mn@doi [\apj] {10.3847/1538-4357/aa6f5d}, \href
  {https://ui.adsabs.harvard.edu/abs/2017ApJ...841...85A} {841, 85}

\bibitem[\protect\citeauthoryear{Alves~Batista, Saveliev, Sigl  \&
  Vachaspati}{Alves~Batista et~al.}{2016}]{AlvesBatista2016}
Alves~Batista R.,  Saveliev A.,  Sigl G.,   Vachaspati T.,  2016, \mn@doi
  [\prd] {10.1103/PhysRevD.94.083005}, 94, 083005

\bibitem[\protect\citeauthoryear{{Berezinskii}, {Bulanov}, {Dogiel}  \&
  {Ptuskin}}{{Berezinskii} et~al.}{1990}]{Berezinskii1990}
{Berezinskii} V.~S.,  {Bulanov} S.~V.,  {Dogiel} V.~A.,   {Ptuskin} V.~S.,
  1990, {Astrophysics of cosmic rays}.
Amsterdam: North-Holland, 1990, edited by Ginzburg, V.L.

\bibitem[\protect\citeauthoryear{{Blasi}, {Amato}  \& {Serpico}}{{Blasi}
  et~al.}{2012}]{blasi2012}
{Blasi} P.,  {Amato} E.,   {Serpico} P.~D.,  2012, \mn@doi [\prl]
  {10.1103/PhysRevLett.109.061101}, \href
  {https://ui.adsabs.harvard.edu/abs/2012PhRvL.109f1101B} {109, 061101}

\bibitem[\protect\citeauthoryear{{Bruno} \& {Carbone}}{{Bruno} \&
  {Carbone}}{2013}]{Bruno2013}
{Bruno} R.,  {Carbone} V.,  2013, \mn@doi [Living Reviews in Solar Physics]
  {10.12942/lrsp-2013-2}, \href
  {https://ui.adsabs.harvard.edu/abs/2013LRSP...10....2B} {10, 2}

\bibitem[\protect\citeauthoryear{Candia \& Roulet}{Candia \&
  Roulet}{2004}]{Candia2004}
Candia J.,  Roulet E.,  2004, \mn@doi [\jcap] {10.1088/1475-7516/2004/10/007},
  0410, 007

\bibitem[\protect\citeauthoryear{Casse, Lemoine  \& Pelletier}{Casse
  et~al.}{2002}]{Casse2001}
Casse F.,  Lemoine M.,   Pelletier G.,  2002, \mn@doi [\prd]
  {10.1103/PhysRevD.65.023002}, 65, 023002

\bibitem[\protect\citeauthoryear{Cesarsky \& Kulsrud}{Cesarsky \&
  Kulsrud}{1973}]{Cesarsky1973}
Cesarsky C.~J.,  Kulsrud R.~M.,  1973, \mn@doi [\apj] {10.1086/152405}, 185,
  153

\bibitem[\protect\citeauthoryear{DeMarco, Blasi  \& Stanev}{DeMarco
  et~al.}{2007}]{DeMarco2007}
DeMarco D.,  Blasi P.,   Stanev T.,  2007, \jcap, 2007, 027

\bibitem[\protect\citeauthoryear{{Evoli}, {Gaggero}, {Grasso}  \&
  {Maccione}}{{Evoli} et~al.}{2008}]{dragon}
{Evoli} C.,  {Gaggero} D.,  {Grasso} D.,   {Maccione} L.,  2008, \mn@doi
  [\jcap] {10.1088/1475-7516/2008/10/018}, \href
  {https://ui.adsabs.harvard.edu/abs/2008JCAP...10..018E} {2008, 018}

\bibitem[\protect\citeauthoryear{{Evoli}, {Aloisio}  \& {Blasi}}{{Evoli}
  et~al.}{2019}]{evoli2019}
{Evoli} C.,  {Aloisio} R.,   {Blasi} P.,  2019, \mn@doi [\prd]
  {10.1103/PhysRevD.99.103023}, \href
  {https://ui.adsabs.harvard.edu/abs/2019PhRvD..99j3023E} {99, 103023}

\bibitem[\protect\citeauthoryear{{Farmer} \& {Goldreich}}{{Farmer} \&
  {Goldreich}}{2004}]{Farmer2004}
{Farmer} A.~J.,  {Goldreich} P.,  2004, \mn@doi [\apj] {10.1086/382040}, \href
  {https://ui.adsabs.harvard.edu/abs/2004ApJ...604..671F} {604, 671}

\bibitem[\protect\citeauthoryear{Fatuzzo, Melia, Todd  \& Adams}{Fatuzzo
  et~al.}{2010}]{Fatuzzo2010}
Fatuzzo M.,  Melia F.,  Todd E.,   Adams F.,  2010, \mn@doi [\apj]
  {10.1088/0004-637X/725/1/515}, 725, 515

\bibitem[\protect\citeauthoryear{{Felice} \& {Kulsrud}}{{Felice} \&
  {Kulsrud}}{2001}]{felice2001}
{Felice} G.~M.,  {Kulsrud} R.~M.,  2001, \mn@doi [\apj] {10.1086/320651}, \href
  {https://ui.adsabs.harvard.edu/abs/2001ApJ...553..198F} {553, 198}

\bibitem[\protect\citeauthoryear{{Gaggero}, {Urbano}, {Valli}  \&
  {Ullio}}{{Gaggero} et~al.}{2015}]{gaggero2015}
{Gaggero} D.,  {Urbano} A.,  {Valli} M.,   {Ullio} P.,  2015, \mn@doi [\prd]
  {10.1103/PhysRevD.91.083012}, \href
  {https://ui.adsabs.harvard.edu/abs/2015PhRvD..91h3012G} {91, 083012}

\bibitem[\protect\citeauthoryear{{Gaggero}, {Grasso}, {Marinelli}, {Taoso}  \&
  {Urbano}}{{Gaggero} et~al.}{2017}]{gaggero_prl2017}
{Gaggero} D.,  {Grasso} D.,  {Marinelli} A.,  {Taoso} M.,   {Urbano} A.,  2017,
  \mn@doi [\prl] {10.1103/PhysRevLett.119.031101}, \href
  {https://ui.adsabs.harvard.edu/abs/2017PhRvL.119c1101G} {119, 031101}

\bibitem[\protect\citeauthoryear{Giacalone \& Jokipii}{Giacalone \&
  Jokipii}{1999}]{Giacalone1999}
Giacalone J.,  Jokipii J.~R.,  1999, \apj, 520, 204

\bibitem[\protect\citeauthoryear{Giacinti, Kachelrie\ss{}, Semikoz  \&
  Sigl}{Giacinti et~al.}{2012}]{Giacinti2012}
Giacinti G.,  Kachelrie\ss{} M.,  Semikoz D.,   Sigl G.,  2012, \jcap, 2012,
  031

\bibitem[\protect\citeauthoryear{Giacinti, Kachelriess  \& Semikoz}{Giacinti
  et~al.}{2018}]{Giacinti2017}
Giacinti G.,  Kachelriess M.,   Semikoz D.~V.,  2018, \mn@doi [\jcap]
  {10.1088/1475-7516/2018/07/051}, 1807, 051

\bibitem[\protect\citeauthoryear{Globus, Allard  \& Parizot}{Globus
  et~al.}{2008}]{Globus2007}
Globus N.,  Allard D.,   Parizot E.,  2008, \mn@doi [\aap]
  {10.1051/0004-6361:20078653}, 479, 97

\bibitem[\protect\citeauthoryear{{Goldstein}}{{Goldstein}}{1976}]{Goldstein1976}
{Goldstein} M.~L.,  1976, \mn@doi [\apj] {10.1086/154239}, \href
  {https://ui.adsabs.harvard.edu/abs/1976ApJ...204..900G} {204, 900}

\bibitem[\protect\citeauthoryear{Grenier}{Grenier}{2019}]{grenier_privcom}
Grenier I.,  2019, private communication

\bibitem[\protect\citeauthoryear{{Grenier}, {Black}  \& {Strong}}{{Grenier}
  et~al.}{2015}]{Grenier2015}
{Grenier} I.~A.,  {Black} J.~H.,   {Strong} A.~W.,  2015, \mn@doi [\araa]
  {10.1146/annurev-astro-082214-122457}, \href
  {https://ui.adsabs.harvard.edu/abs/2015ARA&A..53..199G} {53, 199}

\bibitem[\protect\citeauthoryear{{HESS Collaboration} et~al.,}{{HESS
  Collaboration} et~al.}{2016}]{hess_gc_2016}
{HESS Collaboration} et~al., 2016, \mn@doi [\nat] {10.1038/nature17147}, \href
  {https://ui.adsabs.harvard.edu/abs/2016Natur.531..476H} {531, 476}

\bibitem[\protect\citeauthoryear{Harari, Mollerach  \& Roulet}{Harari
  et~al.}{2014}]{Harari2013}
Harari D.,  Mollerach S.,   Roulet E.,  2014, \mn@doi [\prd]
  {10.1103/PhysRevD.89.123001}, 89, 123001

\bibitem[\protect\citeauthoryear{Harari, Mollerach  \& Roulet}{Harari
  et~al.}{2015}]{Harari2015}
Harari D.,  Mollerach S.,   Roulet E.,  2015, \mn@doi [\prd]
  {10.1103/PhysRevD.92.063014}, 92, 063014

\bibitem[\protect\citeauthoryear{{Hunter}}{{Hunter}}{2007}]{Hunter_2007}
{Hunter} J.~D.,  2007, \mn@doi [Computing in Science Engineering]
  {10.1109/MCSE.2007.55}, 9, 90

\bibitem[\protect\citeauthoryear{Hussein \& Shalchi}{Hussein \&
  Shalchi}{2014}]{Hussein2014}
Hussein M.,  Shalchi A.,  2014, \mn@doi [\apj] {10.1088/0004-637x/785/1/31},
  785, 31

\bibitem[\protect\citeauthoryear{{Iacobelli} et~al.,}{{Iacobelli}
  et~al.}{2013}]{Iacobelli2013}
{Iacobelli} M.,  et~al., 2013, \mn@doi [\aap] {10.1051/0004-6361/201322013},
  \href {https://ui.adsabs.harvard.edu/abs/2013A&A...558A..72I} {558, A72}

\bibitem[\protect\citeauthoryear{{Istomin} \& {Kiselev}}{{Istomin} \&
  {Kiselev}}{2018}]{Istomin2018}
{Istomin} Y.~N.,  {Kiselev} A.~M.,  2018, \mn@doi [\prd]
  {10.1103/PhysRevD.98.083026}, \href
  {https://ui.adsabs.harvard.edu/\#abs/2018PhRvD..98h3026I} {98, 083026}

\bibitem[\protect\citeauthoryear{{Jansson} \& {Farrar}}{{Jansson} \&
  {Farrar}}{2012}]{Jannson2012}
{Jansson} R.,  {Farrar} G.~R.,  2012, \mn@doi [\apj]
  {10.1088/0004-637X/757/1/14}, \href
  {https://ui.adsabs.harvard.edu/abs/2012ApJ...757...14J} {757, 14}

\bibitem[\protect\citeauthoryear{{Jokipii}}{{Jokipii}}{1966}]{Jokipii1966}
{Jokipii} J.~R.,  1966, \mn@doi [\apj] {10.1086/148912}, \href
  {http://adsabs.harvard.edu/abs/1966ApJ...146..480J} {146, 480}

\bibitem[\protect\citeauthoryear{{Jokipii}, {Levy}  \& {Hubbard}}{{Jokipii}
  et~al.}{1977}]{Jokipii1977}
{Jokipii} J.~R.,  {Levy} E.~H.,   {Hubbard} W.~B.,  1977, \mn@doi [\apj]
  {10.1086/155218}, \href {http://adsabs.harvard.edu/abs/1977ApJ...213..861J}
  {213, 861}

\bibitem[\protect\citeauthoryear{{Jones}, {Kaiser}  \& {Birmingham}}{{Jones}
  et~al.}{1973}]{Jones1973}
{Jones} F.~C.,  {Kaiser} T.~B.,   {Birmingham} T.~J.,  1973, International
  Cosmic Ray Conference, \href
  {https://ui.adsabs.harvard.edu/abs/1973ICRC....2..669J} {2, 669}

\bibitem[\protect\citeauthoryear{{Kissmann}}{{Kissmann}}{2014}]{picard}
{Kissmann} R.,  2014, \mn@doi [\apj] {10.1016/j.astropartphys.2014.02.002},
  \href {https://ui.adsabs.harvard.edu/abs/2014APh....55...37K} {55, 37}

\bibitem[\protect\citeauthoryear{{Kleimann}, {Schorlepp}, {Merten}  \& {Becker
  Tjus}}{{Kleimann} et~al.}{2019}]{kleimann2019}
{Kleimann} J.,  {Schorlepp} T.,  {Merten} L.,   {Becker Tjus} J.,  2019,
  \mn@doi [\apj] {10.3847/1538-4357/ab1913}, \href
  {https://ui.adsabs.harvard.edu/abs/2019ApJ...877...76K} {877, 76}

\bibitem[\protect\citeauthoryear{Kluyver et~al.,}{Kluyver
  et~al.}{2016}]{jupyter-notebook}
Kluyver T.,  et~al., 2016, in Loizides F.,  Schmidt B.,  eds, Positioning and
  Power in Academic Publishing: Players, Agents and Agendas. pp 87 -- 90

\bibitem[\protect\citeauthoryear{Kubo}{Kubo}{1957}]{Kubo1957}
Kubo R.,  1957, \mn@doi [Journal of the Physical Society of Japan]
  {10.1143/JPSJ.12.570}, 12, 570

\bibitem[\protect\citeauthoryear{{Kulsrud} \& {Pearce}}{{Kulsrud} \&
  {Pearce}}{1969}]{Kulsrud1969}
{Kulsrud} R.,  {Pearce} W.~P.,  1969, \mn@doi [\apj] {10.1086/149981}, \href
  {https://ui.adsabs.harvard.edu/abs/1969ApJ...156..445K} {156, 445}

\bibitem[\protect\citeauthoryear{{Lange}, {Spanier}, {Battarbee}, {Vainio}  \&
  {Laitinen}}{{Lange} et~al.}{2013}]{lange2013}
{Lange} S.,  {Spanier} F.,  {Battarbee} M.,  {Vainio} R.,   {Laitinen} T.,
  2013, \mn@doi [\aap] {10.1051/0004-6361/201220804}, \href
  {https://ui.adsabs.harvard.edu/abs/2013A&A...553A.129L} {553, A129}

\bibitem[\protect\citeauthoryear{{Mace}, {Dalena}  \& {Matthaeus}}{{Mace}
  et~al.}{2012}]{Mace2012}
{Mace} R.~L.,  {Dalena} S.,   {Matthaeus} W.~H.,  2012, \mn@doi [Phys. Plasmas]
  {10.1063/1.3693379}, \href
  {https://ui.adsabs.harvard.edu/abs/2012PhPl...19c2309M} {19, 032309}

\bibitem[\protect\citeauthoryear{Matthaeus, Qin, Bieber  \& Zank}{Matthaeus
  et~al.}{2003}]{Matthaeus2003}
Matthaeus W.~H.,  Qin G.,  Bieber J.~W.,   Zank G.~P.,  2003, \mn@doi [\apj]
  {10.1086/376613}, 590, L53

\bibitem[\protect\citeauthoryear{McKinney}{McKinney}{2010}]{pandas}
McKinney W.,  2010, in van~der Walt S.,  Millman J.,  eds, Proceedings of the
  9th Python in Science Conference. pp 51 -- 56

\bibitem[\protect\citeauthoryear{{Mertsch}}{{Mertsch}}{2019}]{Mertsch2019}
{Mertsch} P.,  2019, arXiv e-prints, \href
  {https://ui.adsabs.harvard.edu/abs/2019arXiv191001172M} {p. arXiv:1910.01172}

\bibitem[\protect\citeauthoryear{Minnie, Bieber, Matthaeus  \& Burger}{Minnie
  et~al.}{2007}]{Minnie2007}
Minnie J.,  Bieber J.~W.,  Matthaeus W.~H.,   Burger R.~A.,  2007, \apj, 663,
  1049

\bibitem[\protect\citeauthoryear{{Monin} \& {Iaglom}}{{Monin} \&
  {Iaglom}}{1975}]{Monin1975}
{Monin} A.~S.,  {Iaglom} A.~M.,  1975, {Statistical fluid mechanics: Mechanics
  of turbulence. Volume 2 /revised and enlarged edition/}

\bibitem[\protect\citeauthoryear{Parizot}{Parizot}{2004}]{Parizot2004}
Parizot E.,  2004, \mn@doi [Nucl. Phys. Proc. Suppl.]
  {10.1016/j.nuclphysbps.2004.10.034}, 136, 169

\bibitem[\protect\citeauthoryear{{Plotnikov}, {Pelletier}  \&
  {Lemoine}}{{Plotnikov} et~al.}{2011}]{Plotnikov2011}
{Plotnikov} I.,  {Pelletier} G.,   {Lemoine} M.,  2011, \mn@doi [\aap]
  {10.1051/0004-6361/201117182}, \href
  {https://ui.adsabs.harvard.edu/\#abs/2011A&A...532A..68P} {532, A68}

\bibitem[\protect\citeauthoryear{{Pothast}, {Gaggero}, {Storm}  \&
  {Weniger}}{{Pothast} et~al.}{2018}]{pothast2018}
{Pothast} M.,  {Gaggero} D.,  {Storm} E.,   {Weniger} C.,  2018, \mn@doi
  [\jcap] {10.1088/1475-7516/2018/10/045}, \href
  {https://ui.adsabs.harvard.edu/abs/2018JCAP...10..045P} {2018, 045}

\bibitem[\protect\citeauthoryear{Qin, Zhang, Xiao, Liu, Sun  \& Tang}{Qin
  et~al.}{2013}]{Qin2013}
Qin H.,  Zhang S.,  Xiao J.,  Liu J.,  Sun Y.,   Tang W.~M.,  2013, \mn@doi
  [Phys. Plasmas] {10.1063/1.4818428}, 20, 084503

\bibitem[\protect\citeauthoryear{{Schlegel}, {Frie}, {Eichmann}, {Reichherzer}
  \& {Tjus}}{{Schlegel} et~al.}{2020}]{Schlegel2019}
{Schlegel} L.,  {Frie} A.,  {Eichmann} B.,  {Reichherzer} P.,   {Tjus} J.~B.,
  2020, \mn@doi [\apj] {10.3847/1538-4357/ab643b}, \href
  {https://ui.adsabs.harvard.edu/abs/2020ApJ...889..123S} {889, 123}

\bibitem[\protect\citeauthoryear{{Schlickeiser}}{{Schlickeiser}}{1989}]{Schlickeiser1989}
{Schlickeiser} R.,  1989, \mn@doi [\apj] {10.1086/167009}, \href
  {https://ui.adsabs.harvard.edu/abs/1989ApJ...336..243S} {336, 243}

\bibitem[\protect\citeauthoryear{{Schlickeiser}}{{Schlickeiser}}{2002}]{Schlickeiser2002}
{Schlickeiser} R.,  2002, Springer-Verlag, Berlin, \href
  {http://adsabs.harvard.edu/abs/2002cra..book.....S} {}

\bibitem[\protect\citeauthoryear{{Schlickeiser}}{{Schlickeiser}}{2015}]{2015PhPl...22i1502S}
{Schlickeiser} R.,  2015, \mn@doi [Physics of Plasmas] {10.1063/1.4928940},
  \href {https://ui.adsabs.harvard.edu/abs/2015PhPl...22i1502S} {22, 091502}

\bibitem[\protect\citeauthoryear{{Schlickeiser}, {Caglar}  \&
  {Lazarian}}{{Schlickeiser} et~al.}{2016}]{schlickeiser2016}
{Schlickeiser} R.,  {Caglar} M.,   {Lazarian} A.,  2016, \mn@doi [\apj]
  {10.3847/0004-637X/824/2/89}, \href
  {https://ui.adsabs.harvard.edu/abs/2016ApJ...824...89S} {824, 89}

\bibitem[\protect\citeauthoryear{{Seta}, {Shukurov}, {Wood}, {Bushby}  \&
  {Snodin}}{{Seta} et~al.}{2018}]{seta2018}
{Seta} A.,  {Shukurov} A.,  {Wood} T.~S.,  {Bushby} P.~J.,   {Snodin} A.~P.,
  2018, \mn@doi [\mnras] {10.1093/mnras/stx2606}, \href
  {https://ui.adsabs.harvard.edu/abs/2018MNRAS.473.4544S} {473, 4544}

\bibitem[\protect\citeauthoryear{Shalchi}{Shalchi}{2009}]{Shalchi2009}
Shalchi A.,  2009, Springer-Verlag, Berlin, Heidelberg, 362

\bibitem[\protect\citeauthoryear{{Shalchi}, {Bieber}, {Matthaeus}  \&
  {Qin}}{{Shalchi} et~al.}{2004}]{Shalchi2004}
{Shalchi} A.,  {Bieber} J.~W.,  {Matthaeus} W.~H.,   {Qin} G.,  2004, \mn@doi
  [\apj] {10.1086/424839}, \href
  {https://ui.adsabs.harvard.edu/abs/2004ApJ...616..617S} {616, 617}

\bibitem[\protect\citeauthoryear{{Shalchi}, {Skoda, T.}, {Tautz, R. C.}  \&
  {Schlickeiser, R.}}{{Shalchi} et~al.}{2009}]{Shalchi2009AA}
{Shalchi} {Skoda, T.} {Tautz, R. C.}  {Schlickeiser, R.} 2009, \mn@doi [\aap]
  {10.1051/0004-6361/200912755}, 507, 589

\bibitem[\protect\citeauthoryear{{Shukurov}, {Rodrigues}, {Bushby}, {Hollins}
  \& {Rachen}}{{Shukurov} et~al.}{2019}]{shukurov2019}
{Shukurov} A.,  {Rodrigues} L. F.~S.,  {Bushby} P.~J.,  {Hollins} J.,
  {Rachen} J.~P.,  2019, \mn@doi [\aap] {10.1051/0004-6361/201834642}, \href
  {https://ui.adsabs.harvard.edu/abs/2019A&A...623A.113S} {623, A113}

\bibitem[\protect\citeauthoryear{Snodin, Shukurov, Sarson, Bushby  \&
  Rodrigues}{Snodin et~al.}{2016}]{Snodin2015}
Snodin A.~P.,  Shukurov A.,  Sarson G.~R.,  Bushby P.~J.,   Rodrigues L. F.~S.,
   2016, \mn@doi [\mnras] {10.1093/mnras/stw217}, 457, 3975

\bibitem[\protect\citeauthoryear{{Sridhar} \& {Goldreich}}{{Sridhar} \&
  {Goldreich}}{1994}]{Sridhar1994}
{Sridhar} S.,  {Goldreich} P.,  1994, \mn@doi [\apj] {10.1086/174600}, \href
  {https://ui.adsabs.harvard.edu/abs/1994ApJ...432..612S} {432, 612}

\bibitem[\protect\citeauthoryear{{Srinivasan} \& {Shalchi}}{{Srinivasan} \&
  {Shalchi}}{2014}]{Srinivasan2014}
{Srinivasan} S.,  {Shalchi} A.,  2014, \mn@doi [\apss]
  {10.1007/s10509-013-1705-x}, \href
  {http://adsabs.harvard.edu/abs/2014Ap%26SS.350..197S} {350, 197}

\bibitem[\protect\citeauthoryear{{Strong} \& {Moskalenko}}{{Strong} \&
  {Moskalenko}}{1998}]{galprop}
{Strong} A.~W.,  {Moskalenko} I.~V.,  1998, \mn@doi [\apj] {10.1086/306470},
  \href {https://ui.adsabs.harvard.edu/abs/1998ApJ...509..212S} {509, 212}

\bibitem[\protect\citeauthoryear{Subedi et~al.,}{Subedi
  et~al.}{2017}]{Subedi2017}
Subedi P.,  et~al., 2017, \apj, 837, 140

\bibitem[\protect\citeauthoryear{{Tautz} \& {Shalchi}}{{Tautz} \&
  {Shalchi}}{2010}]{Tautz2010}
{Tautz} R.~C.,  {Shalchi} A.,  2010, \mn@doi [Phys. Plasmas]
  {10.1063/1.3530185}, \href
  {https://ui.adsabs.harvard.edu/abs/2010PhPl...17l2313T} {17, 122313}

\bibitem[\protect\citeauthoryear{Tautz, Shalchi  \& Schlickeiser}{Tautz
  et~al.}{2008}]{Tautz2008}
Tautz R.~C.,  Shalchi A.,   Schlickeiser R.,  2008, \mn@doi [\apj]
  {10.1086/592498}, 685, L165

\bibitem[\protect\citeauthoryear{{Virtanen} et~al.,}{{Virtanen}
  et~al.}{2019}]{Virtanen2019}
{Virtanen} P.,  et~al., 2019, arXiv e-prints, \href
  {https://ui.adsabs.harvard.edu/abs/2019arXiv190710121V} {p. arXiv:1907.10121}

\bibitem[\protect\citeauthoryear{V{\"o}lk}{V{\"o}lk}{1973}]{Voelk1973}
V{\"o}lk H.~J.,  1973, \mn@doi [\apss] {10.1007/BF00649186}, 25, 471

\bibitem[\protect\citeauthoryear{Winkel, Speck  \& Ruprecht}{Winkel
  et~al.}{2015}]{Winkel2015}
Winkel M.,  Speck R.,   Ruprecht D.,  2015, \mn@doi [PAMM]
  {10.1002/pamm.201510333}, 15, 687

\bibitem[\protect\citeauthoryear{Yan \& Lazarian}{Yan \&
  Lazarian}{2008}]{Yan2008}
Yan H.,  Lazarian A.,  2008, \mn@doi [\apj] {10.1086/524771}, 673, 942

\bibitem[\protect\citeauthoryear{{Yang}, {Aharonian}  \& {Evoli}}{{Yang}
  et~al.}{2016}]{yang2016}
{Yang} R.,  {Aharonian} F.,   {Evoli} C.,  2016, \mn@doi [\prd]
  {10.1103/PhysRevD.93.123007}, \href
  {https://ui.adsabs.harvard.edu/abs/2016PhRvD..93l3007Y} {93, 123007}

\bibitem[\protect\citeauthoryear{{Zank}, {Matthaeus}  \& {Smith}}{{Zank}
  et~al.}{1996}]{Zank1996}
{Zank} G.~P.,  {Matthaeus} W.~H.,   {Smith} C.~W.,  1996, \mn@doi [\jgr]
  {10.1029/96JA01275}, \href
  {https://ui.adsabs.harvard.edu/abs/1996JGR...10117093Z} {101, 17093}

\bibitem[\protect\citeauthoryear{Zweibel}{Zweibel}{2013}]{Zweibel2013}
Zweibel E.~G.,  2013, \mn@doi [Phys. Plasmas] {10.1063/1.4807033}, 20, 055501

\bibitem[\protect\citeauthoryear{{Zweibel}}{{Zweibel}}{2017}]{2017PhPl...24e5402Z}
{Zweibel} E.~G.,  2017, \mn@doi [Physics of Plasmas] {10.1063/1.4984017}, \href
  {https://ui.adsabs.harvard.edu/abs/2017PhPl...24e5402Z} {24, 055402}

\bibitem[\protect\citeauthoryear{{van der Walt}, {Colbert}  \&
  {Varoquaux}}{{van der Walt} et~al.}{2011}]{van_der_Walt_2011}
{van der Walt} S.,  {Colbert} S.~C.,   {Varoquaux} G.,  2011, \mn@doi
  [Computing in Science Engineering] {10.1109/MCSE.2011.37}, 13, 22

\makeatother
\end{thebibliography}



\appendix
\section{Decorrelated Particle Trajectories}\label{app:A}
\begin{figure}
	\includegraphics[width=\columnwidth]{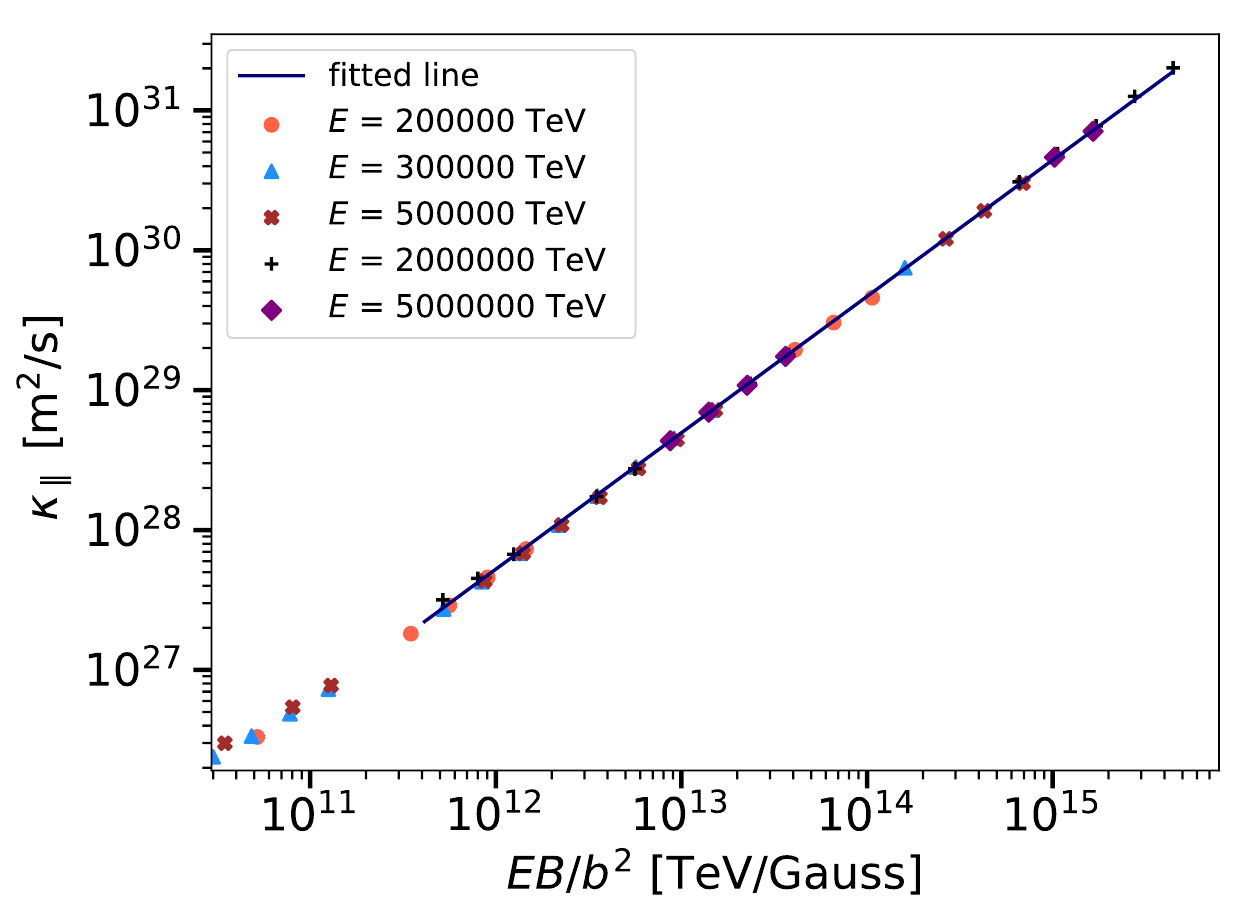}\caption{Parallel diffusion coefficient as functions of the ratio $EB/b^2$. The presented fit confirms
the predicted dependency of the parallel diffusion coefficient $\kappa_\parallel \propto EB/b^2$ (see Eq.~(\ref{D_final})) for simulations with step sizes $l \lesssim s$ that are on the order of the magnetic fluctuations. {{The shown data points meet this condition, because the step sizes for 200~PeV are already $1.3\, l_\mathrm{c}$}}. The step size scales linearly with the particle energy according to $s = r_\mathrm{g}/10$.
The slope of the presented fit reads $0.975 \pm 0.004$.}
	\label{fig:high_energy_pollution}
\end{figure}

For numerical simulations, the step size needs to resolve the gyromotion and the scale of the magnetic fluctuations. The latter condition requires many steps per gyration for high-energy particles and thus long simulation times. In the case of step sizes that are larger than the scale of the fluctuations, the turbulent magnetic field vectors can be assumed to be deccorrelated between two subsequent particle positions along the particle trajectory. Without the assumption of correlated turbulence, the turbulent magnetic field $\textbf{b}$ at an arbitrary position points into a random direction, so that $b_{i}$ is also arbitrary: $-b \leq b_{i} \leq b$. 
The key aspect is, however, {{that the root-mean-square value of $b_i$ is proportional to the root-mean-square value of $b$}}.
It is now possible to pull the magnetic field from Eq.~(\ref{Diff_Momentum}) in front of the integral as shown below:
\begin{equation}
\begin{split}
D_{33}(v) = b^2\left(\frac{q}{m\gamma}\right)^2 \int \limits_0^t &\mathrm{d}\tau~ \frac{1}{2\pi}\int \limits_0^{2\pi}\mathrm{d}\theta ~(  v_{{1}}(0,\theta) v_{{1}} (\tau,\theta)+  \\
v_{{2}}(0,\theta) v_{{2}} (\tau,\theta) &-v_{{2}}(0,\theta) v_{{1}} (\tau,\theta)  -  v_{{1}}(0,\theta) v_{{2}} (\tau,\theta) 
).
\end{split}
\end{equation}
Here, the parallel component of the diffusion tensor is considered.\\\\
For small ratios $b/B$, particles follow a helical trajectory caused by the background magnetic field in the $x_3$-direction. This motion can be separated into the motion of the gyrocenter with a position $\textbf{X}$ and the circular motion along the trajectory $\textbf{s}$
\begin{align}\label{def_v}
\textbf{s} = \left( \begin{array}{c}
\sin(\theta) \\
-\cos(\theta)\\
0
\end{array} \right)~;~\textbf{X} = \left( \begin{array}{c}
0\\0\\
v_{\parallel}(0)\tau
\end{array} \right),
\end{align}
which orders out drift velocities.
The velocity $\textbf{v}$ of a gyrating particle can therefore be parameterised together with its positions as
\begin{align}
\textbf{v} = \left( \begin{array}{c}
v_{\perp}\cos(\theta) \\
v_{\perp}\sin(\theta) \\
v_{\parallel}
\end{array} \right)=v_\parallel \textbf{e}_3 +v_\perp \textbf{c}(\theta)~;~\textbf{c} = \left( \begin{array}{c}
\cos(\theta) \\
\sin(\theta) \\
0
\end{array} \right).
\end{align}
Using this parameterisation for the particle velocity results in
\begin{equation}\label{xx}
\begin{split}
D_{33}(v) &= \left(\frac{qbv_\perp}{m\gamma }\right)^2 \int \limits_0^t \mathrm{d}\tau~\left(\cos\left(\frac{v_{\perp}}{r_g}\tau\right)-\sin\left(\frac{v_{\perp}}{r_g}\tau\right)\right)~\\ &\frac{1}{2\pi}\int \limits_0^{2\pi}\mathrm{d}\theta ~(\cos^2\theta~+\sin^2\theta),\\
&~~~= \left(\frac{qbv_\perp}{m\gamma}\right)^2 \int \limits_0^t \mathrm{d}\tau~\left(\cos\left(\frac{v_{\perp}}{r_g}\tau\right)-\sin\left(\frac{v_{\perp}}{r_g}\tau\right)\right).
\end{split}
\end{equation}
Substituting $\hat{\tau} = \tau v_{\perp}/r_{\mathrm{g}}$ together with d$\tau =  r_{\mathrm{g}}/v_{\perp}$ d$\hat{\tau}$ results in
\begin{align}
D_{33}(v) \propto \left(\frac{qbv_{\perp}}{m\gamma }\right)^2  ~\frac{r_{\mathrm{g}}}{v_{\perp}}~ \int \limits_0^{t'} \mathrm{d}\hat{\tau}~ ( \cos \hat{\tau} -  \sin \hat{\tau}
).
\end{align}
With this parallel momentum diffusion coefficient, it is possible to derive the parallel spatial diffusion coefficient based on Eq.$~$(\ref{kappa D}) as
\begin{align}\label{D_final}
\kappa_{\parallel} \propto \left(\frac{m\gamma  v^2}{qbv_{\perp}}\right)^2  ~\frac{v_{\perp}}{r_{\mathrm{g}}} \propto \frac{m^2\gamma^2v^4}{b^2 v_{\perp}r_{\mathrm{g}}}\propto  \frac{B}{ b^2}\frac{E}{q} \propto \left(\frac{B}{b}\right)^2  r_{\mathrm{g}}c.
\end{align}
Agreement between simulated data for high-energy particles and this relation can be seen in Fig.~\ref{fig:high_energy_pollution}, where the parallel diffusion coefficient is shown as a function of the right-hand side of Eq.~(\ref{D_final}).

\section{Scaling of the results with reduced rigidity}\label{app:B}
Given the prediction of QLT, $\kappa_\parallel = cl_\mathrm{c}(r_\mathrm{g}/l_\mathrm{c})^\gamma (B/b)^2$
makes the problem rescalable for a different range of energies $E$, magnetic field properties $b,B,l_\mathrm{c}$ and the particle's electric charge $q$:
\begin{equation}
\begin{split}
    \kappa_\parallel = 9.494\cdot 10^{27} \mathrm{cm}^2/\mathrm{s}\,\cdot 0.9251^{-\gamma +1/3}\cdot \left(\frac{b}{0.1\, \mathrm{\mu G}}\right)^{-2}\\ \left(\frac{B}{\mathrm{\mu G}}\right)^{2-\gamma} \left(\frac{E}{10\, \mathrm{PeV}}\right)^{-2} \left(\frac{l_\mathrm{c}}{10\, \mathrm{pc}}\right)^{-2} \left(\frac{q}{\mathrm{e}}\right)^{-2}.
    \label{kpar:equ}
    \end{split}
\end{equation}
The boundaries of the RSR derived within this study can be rescaled analogously as follows. Combining the expressions for the gyroradius of highly relativistic particles and the definition of the lower boundary of the RSR,
\begin{equation*}
    r_\mathrm{g} = \frac{E}{cqB} = \frac{l_\mathrm{min}}{\pi (b/B)},
\end{equation*}
results in the lower-limit energy of the RSR
\begin{equation}
E_\mathrm{min} = 294.5\,\mathrm{PeV}\,  \left(\frac{b}{\mathrm{\mu G}}\right)^{-1} \left(\frac{B}{\mathrm{\mu G}}\right)^2 \left(\frac{l_\mathrm{min}}{\mathrm{pc}}\right) \left(\frac{q}{\mathrm{e}}\right).
\end{equation}
The maximum energy of particles in the RSR yields
\begin{equation}
E_\mathrm{max} = 14.72\,\mathrm{PeV}\,  \left(\frac{B}{\mathrm{\mu G}}\right) \left(\frac{l_\mathrm{max}}{\mathrm{pc}}\right) \left(\frac{q}{\mathrm{e}}\right).
\end{equation}
The diffusion process of particles is consequently not limited to a certain range of energies, but can be rescaled accordingly, assuming that (a) the power-law behaviour can be extended to the entire range of the turbulence spectrum $(l_{\min},\,l_{\max})$ and (b) assuming that the $(b/B)$-dependence is as expected in Eq.\ (\ref{kpar:equ}). We find evidence in our simulations that the $(b/B)^{-2}-$ dependence holds (Reichherzer et al., in prep).
Therefore, scaling can be considered applicable in several astrophysical environments as demonstrated with two examples of cosmic-ray propagation sites:
\begin{enumerate}
    \item \textit{Heliosphere}: Typical magnetic field parameters at 1 AU in the heliosphere are $b/B \approx 0.4-1$ \citep{Bruno2013}, $B \sim 5\,$nT \citep{Giacalone1999,Adhikari2017} as well as $l_\mathrm{min} \sim 0.1\,$AU and $l_\mathrm{max} \sim 30\,$AU \citep{Zank1996}. Here, the RSR for protons lies within $3-7\,\mathrm{GeV} \lesssim E \lesssim 107\,\mathrm{TeV}$. Care must be taken with this lower limit, since the protons around $E_\mathrm{min}$ can hardly be treated as highly relativistic, a property which is used in the calculations in this paper to make the analysis feasible.
    \item \textit{Galaxy}: The magnetic waves of the turbulence range between the dissipation scale $l_\mathrm{min} \sim 1\,\mathrm{AU} $ and the maximum scale $l_\mathrm{max} \sim 150\,\mathrm{pc}$ in the halo and $l_\mathrm{max} \sim 20\,\mathrm{pc}$ in the disk \citep{Iacobelli2013}. Assuming $B\, \sim \mathrm{\mu G}$ and $b\, \sim 0.1\,\mathrm{\mu G}$ constrains the RSR within $14\,\mathrm{GeV} \lesssim E_\mathrm{halo} \lesssim 22\,\mathrm{PeV}$ and $14\,\mathrm{GeV} \lesssim E_\mathrm{disk} \lesssim 3\,\mathrm{PeV}$ for protons.
\end{enumerate}

\bsp	
\label{lastpage}

\end{document}